\documentclass[english,onecolumn,prx,preprintnumbers,amsmath,amssymb,superscriptaddress]{revtex4-2}
\usepackage{verbatim}
\usepackage{graphicx}
\usepackage{amssymb}
\usepackage{ulem}
\usepackage{xcolor}
\usepackage{soul}
\usepackage{hyperref}

\usepackage{babel}
\usepackage{float}
\usepackage{comment}
\begin{document}
%	\setcitestyle{super}
	\preprint{XXX}

	\title{Observation of tunable discrete time crystalline phases}
	
	\author{Arnab Sarkar}
	\affiliation{Department of Physics, Indian Institute of Technology - Kanpur, UP-208016, India}
	\author{Anurag}
	\affiliation{Department of Physics, Indian Institute of Technology - Kanpur, UP-208016, India}
	\author{Javed A. Mondal}
	\affiliation{Department of Physics, Indian Institute of Technology - Kanpur, UP-208016, India}
	\author{Rajan Singh}
	\affiliation{Department of Physics, Indian Institute of Technology - Kanpur, UP-208016, India}
	%\affiliation{Department of Condensed Matter Physics, Weizmann Institute of Science -234 Herzl Street, POB 26,Rehovot-7610001, Israel}
	\author{Aamir A. Makki}
	\affiliation{Department of Physics, Indian Institute of Technology - Kanpur, UP-208016, India}
	\author{Ateesh K. Rathi}
	\affiliation{Department of Physics, Indian Institute of Technology - Kanpur, UP-208016, India}
	
	\author{Ryan J.T. Nicholl}
	\affiliation{Department of Physics and Astronomy, Vanderbilt University, Nashville, Tennessee 37235, USA}	
	
	\author{Sagar Chakraborty}
	\affiliation{Department of Physics, Indian Institute of Technology - Kanpur, UP-208016, India}
	
	\author{Kirill I. Bolotin}
	\affiliation{Department of Physics, Freie Universitat Berlin, Arnimallee 14, Berlin 14195, Germany}
	
	\author{Saikat Ghosh}
	\email{gsaikat@iitk.ac.in}
	\affiliation{Department of Physics, Indian Institute of Technology - Kanpur, UP-208016, India}

	\date{\today}
	
	\begin{abstract}
        Discrete time crystals (DTCs) are non-equilibrium phases of matter that emerge due to the breaking of time translational symmetry of periodically driven many-body systems. A non-ergodic, rigid subharmonic response of a many-body system is a typical signature of the existence of a time crystal. Here, we report experimental observation of multiple novel DTC phases, including a subharmonic response, in a classical nanoelectromechanical system (NEMS) based on coupled graphene and silicon nitride membranes. Unlike prethermal-time crystalline phases of a closed system, stability of the phases is ensured by the presence of linear and nonlinear damping in the system, and the phases are stable in the true sense. Furthermore, we employ controlled mechanical strain to drive the transitions between different phases and show a phase diagram. Finally we model the system theoretically using a mean-field description to study and compare the system in parameter space.
	\end{abstract}

	\maketitle
	
	%\section{Introduction}
\paragraph*{\textbf{Introduction:}}    
Over last decade, renewed interest in possible temporal symmetry breaking resulted in the discovery of discrete time-crystals (DTCs). These emergent phases of interacting periodically driven systems~\cite{PhysRevLett.109.160402, PhysRevLett.117.090402,PhysRevLett.116.250401,PhysRevLett.118.030401,doi:10.1063/PT.3.4019,Mi2022,doi:10.1126/science.abg8102,doi:10.1126/science.abg2530,PhysRevLett.127.090602} have been suggested for applications in computation, metrology, and sensing ~\cite{doi:10.1146/annurev-conmatphys-031119-050658,https://doi.org/10.48550/arxiv.1910.10745,Sacha_2017}.  In a DTC phase, a periodically driven many-body system settles into a common period that is longer than the drive, breaking the discrete time translational symmetry. Such collective breaking of discrete time translational symmetry of the periodic drive results in a non-ergodic dynamical steady state, the DTC phase, that exhibits novel features reminiscent of equilibrium thermodynamic phases. In a closed system, prethermal DTC phases are observed, which is stabilized by localization, disorder, and many-body interaction but eventually thermalizes after a very long time ~\cite{Zhang2017,Choi2017,doi:10.1126/science.abg8102}. For an open system, DTC phases are stabilized by the interplay between periodic drive, nonlinearity, linear and nonlinear damping and the many body nature of the system~\cite{PhysRevLett.126.020603,Yao2020,PhysRevLett.127.043602,pub.1117345355}.

Recently, many experiments have been performed in quantum systems with ions, atoms and defects in solids to observe signatures of prethermal DTC in the form of subharmonic entrainment~\cite{Choi2017,Zhang2017,Autti2022}. In classical dynamical systems, such discrete time translation symmetry breaking is ubiquitous, ranging from parametrically driven swings to Faraday waves~\cite{pub.1026673574}. In presence of fluctuation, subharmonic oscillation of a classical harmonic oscillator fades away but a many body coupled oscillators shows robust subharmonic response which is identified as a DTC phase ~\cite{PhysRevLett.126.020603,https://doi.org/10.48550/arxiv.1910.10745, doi:10.1146/annurev-conmatphys-031119-050658, doi:10.1063/PT.3.4020, https://doi.org/10.48550/arxiv.2203.05577}. Even with two weakly coupled parametric oscillators, common overlap of dynamical instability regions of the resonator modes has led to DTC-like features~\cite{PhysRevLett.123.124301}. 
    
Nanoelectromechanical systems (NEMS) are great platform to study DTC phases because of the presence of giant nonlinearity, nonlinear damping~\cite{Singh2019Giant}, parameter tunability ~\cite{Singh2018} and low actuation power. In this letter we study a NEMS device in which multiple mechanical modes of a Silicon Nitride (SiNx) membrane resonator simultaneously interact with two spatially separated graphene resonators. Exceptional elastic and mechanical non-linear properties of graphene~\cite{RevModPhys.94.045005} make the interaction tunable, resulting in long-range interaction between otherwise non-interacting many (on average 16) SiNx modes (see SI, section III. A). When driven parametrically at twice of the average frequency of the graphene-SiNx hybrid modes, universal collective modes emerge above a threshold drive power, breaking discrete time translational symmetry of the drive. Remarkably, we observe a rich palette of distinct DTC phases:  sub-harmonic, anharmonic and a novel biharmonic phase~\cite{Zhang2017,Choi2017,Autti2022,doi:10.1126/sciadv.aay8892,PhysRevResearch.3.023106,PhysRevLett.123.083901}. Furthermore, the mean field analysis confirms the origin of three new phases through bifurcations of fixed points and limit cycles~\cite{PhysRevLett.117.214101,PhysRevLett.123.083901,https://doi.org/10.48550/arxiv.2203.05577}.

	\begin{figure*}
		\centering
		\includegraphics[scale=0.21]{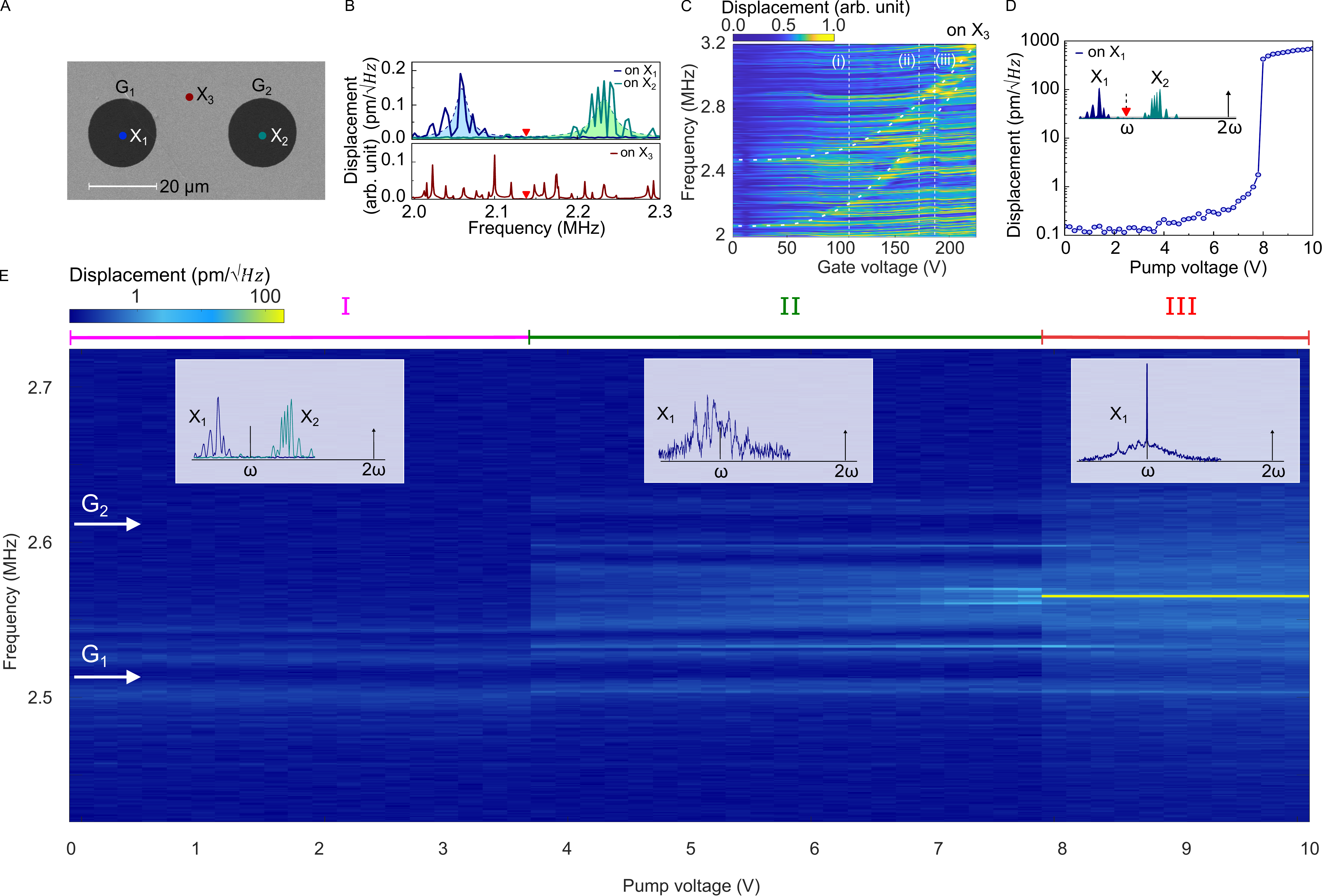}
		\caption{\textbf{Graphene/SiNx NEMS supporting DTC phases:} \textbf{(A)} SEM image of the device comprised of two graphene drums (black, the two $20$ $\mu m$ diameter drums are used for all experiments) suspended on a SiNx membrane (gray). Three dots on the sample depict the detection points on graphene ($X_1$, $X_2$) and SiNx ($X_3$) surface (blue, green and red, respectively). \textbf{(B)} Blue and green curves (upper panel) show thermal spectra of the two graphene drums ($G_1$ and $G_2$) measured at $X_1$ and $X_2$. The spectrum is described by two Lorentzian peaks due to graphene drums (dotted curves) hybridized with multiple SiNx modes. These modes are visible directly in a measurement (at $X_3$) on SiNx surface (lower panel). \textbf{(C)} Weakly driven spectra of SiNx, measured at $X_3$. With varying DC gate voltage (controlling the tension), we observe dispersion of the two graphene modes with tunable frequency separation (white dotted curves). These modes hybridize with different sets of SiNx modes at different frequencies. Three typical phases of Figs.~2A, B and C, are observed at voltages corresponding to (ii), (i) and (iii), respectively. \textbf{(D)} Spectral amplitude at the midpoint ($\omega$, marked by red arrow in inset) of the two resonators, with increasing amplitude of a parametric drive (AC gate voltage) at a frequency $2 \omega$. \textbf{(E)} Displacement spectra measured at $X_1$, corresponding to DC gate voltage (ii) of Fig.~1C, with increasing amplitude of the above-mentioned parametric drive. Region I exhibits the isolated cluster of hybrid modes corresponding to $G_1$, region II shows a sharp transition to a correlated liquid-like phase while in region III, the system undergoes a transition to a DTC phase.} 
      \label{device}
	\end{figure*}

	%\section{Device and experiment}
\paragraph*{\textbf{Device and experiment:}} 

The sample consists of a  $320 \times 320$ $\mu m^2$ SiNx resonator with graphene monolayer deposited onto two $20$ $\mu$m holes drilled in its surface, forming a second set of resonators  $G_1$ and $G_2$ (Fig.~\ref{device}A). It is arranged in a capacitive configuration, with a highly doped silicon surface used as a backgate electrode.  The gate voltage applied across the device is used to tune and actuate the vibrational modes. The device is placed in a vacuum chamber at room temperature for measurement. A custom-built confocal microscope, integrated with a Michelson interferometer and active feedback control, is employed to study the device. Vibrational spectra are measured by focusing the external  cavity diode laser (ECDL) on either of the graphene drums ($X_1$ or $X_2$ in Fig.~\ref{device}A) or SiNx ($X_3$ in Fig.~\ref{device}A)~\cite{Singh2018}. When measured on SiNx, a low inbuilt tension and high quality factor result in individually resolvable but densely packed spectra of modes (Fig.~\ref{device}B, below). In contrast, measurements on graphene reveal lower quality factors fundamental modes of the relevant drums hybridized with multiple SiNx modes (Fig.~\ref{device}B, above). This hybridization induces effective nonlinear cross-talk between the SiNx modes (Fig.~\ref{device}B)~\cite{Singh2019Giant}. As the system tension is increased, via DC gate voltage, the frequencies of flexible graphene drums shift upwards (dotted lines in Fig.~\ref{device}C) while the modes of rigid SiNx remain virtually unchanged (horizontal lines in Fig.~\ref{device}C). Nevertheless, spectra of $G_1$ and $G_2$ remain decoupled in the entire range of DC gate voltages and interact with differing sets of SiNx modes, with no signature of direct interaction (e.g. avoid crossing vs. DC gate voltage~\cite{Singh2019Giant}) between the two graphene resonators is observed.

	\begin{figure}
		\centering
		\includegraphics[scale=0.13]{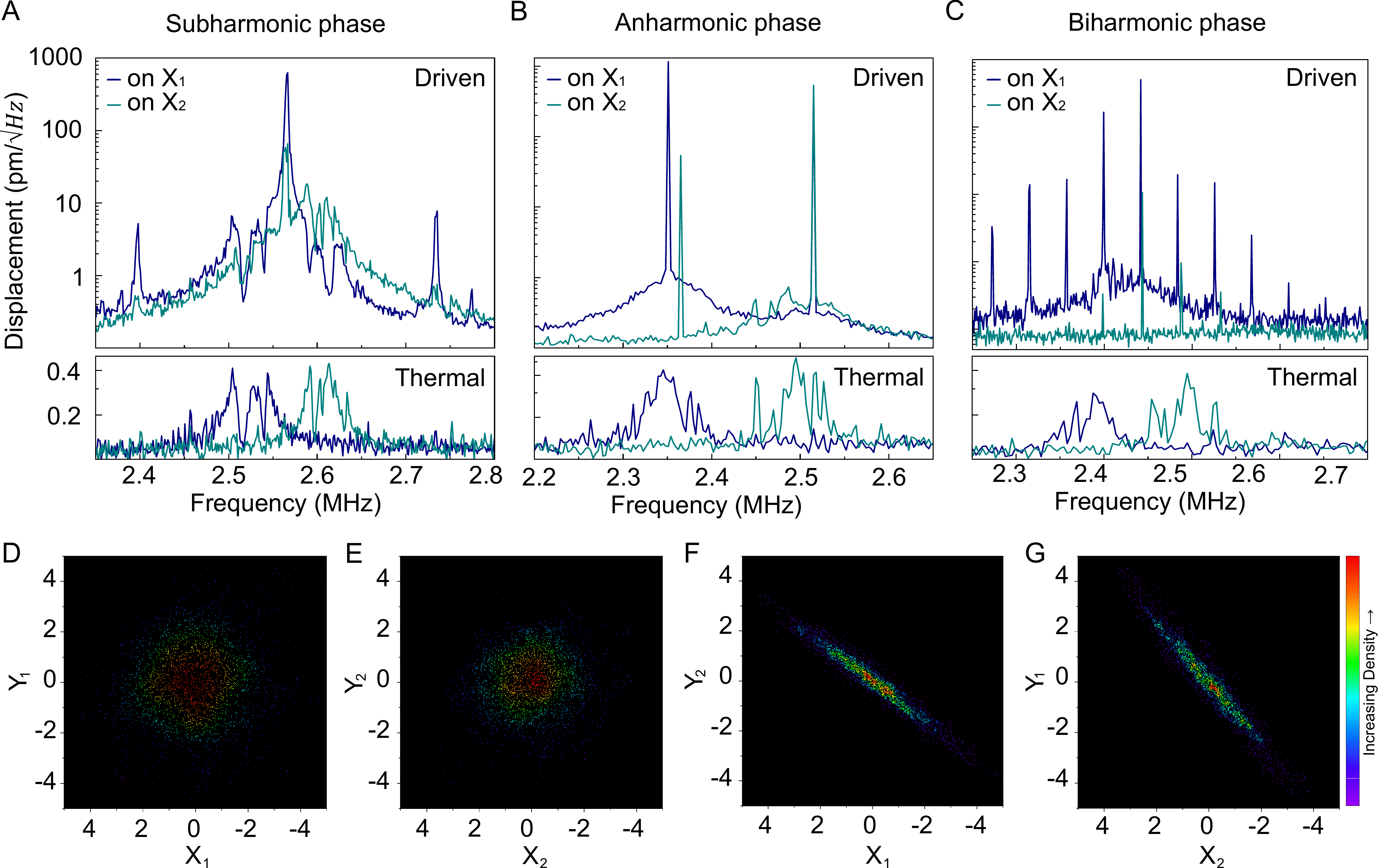}
		\caption{\textbf{Emergent time-crystalline phases at high drive strength:}   \textbf{(A)} Subharmonic DTC phase observed at gate voltage (ii), Fig.~1C. \textbf{(B)} Anharmonic DTC phase with gate voltage (i), Fig.~1C. \textbf{(D)} Biharmonic DTC phase with gate voltage (iii), Fig.~1C. Lower panels of A-C show corresponding thermal peaks. All measurements carried out at the point $X_1$ (blue) and $X_2$ (green).\textbf{(D, E)} Quadrature plots ($X_1 - Y_1$ and $X_2 - Y_2$) corresponding to the two peaks  of the anharmonic phase (Fig.~2B) show thermal-like fluctuations. \textbf{(F, G)} Cross quadratures ($X_1 - Y_2$ and $X_2 - Y_1$) show strong anti-correlation.} 
        \label{DTC_phases_quad}
	\end{figure}

	%\section{Emergent DTC phases}
\paragraph*{\textbf{Observation of DTC phases:}} 
 To access time-crystalline behaviours, the device is parametrically driven at twice of the mean frequency of the clusters of hybrid modes around G1 and G2 (inset, Fig.~\ref{device}D). With increasing drive strength, the spectral amplitude at $\omega$ first rises from the noise floor at $4$V and then undergoes a sharp transition to a high-amplitude steady state at around $8$V (Fig.~\ref{device}E). In region I of Fig.~\ref{device}E, the corresponding spectra contain distinct signatures of graphene-SiNx interactions of the separated clusters of thermal modes around G1 and G2. With increasing drive voltage, the two clusters merge sharply (around 4V, region II) to a single shared spectra, that nevertheless remain thermal-like with small amplitude and without any sharp oscillation frequencies.  Eventually, at around 8V, a sharp subharmonic peak, emerges at frequency $\omega$ with oscillation amplitude that are orders of magnitude higher compared to the thermal modes (region III). The DTC spectra, integrated over millions of oscillation periods and measured on resonator surfaces at $X_1$, $X_2$ (blue and green respectively, in Fig.~\ref{device}A) and $X_3$,  show evidence for emergence of long-range order in time and space due to interactions between modes of spatially distinct graphene and SiNx oscillators, in contrast to the local features observed below the threshold (Fig.~\ref{device}B).

    Tuning the frequency separation of graphene and sets of SiNx with DC gate voltage, we observe two other distinct  phases with similar long-range order properties in addition to the \textit{subharmonic phase} shown in Fig.~\ref{DTC_phases_quad}A. These are  \textit{anharmonic} (Fig.~\ref{DTC_phases_quad}B) and \textit{biharmonic} (Fig.~\ref{DTC_phases_quad}C) phases, distinguished by even or odd numbers of frequency comb lines in the spectra. Indeed, several recent theoretical works predict the first two of these DTC phases~\cite{PhysRevResearch.3.023106,PhysRevLett.118.254301,PhysRevLett.123.083901}. To the best of our knowledge, the ~\textit{biharmonic} phase is novel. In the remainder of the manuscript, we confirm that assignment of these as time-crystalline phases, map their phase diagram, and explore transitions between them. 
	
 It can be noted that the anharmonic DTC phase provides an indirect means of quantifying correlations by noting that the two sharp spectral peaks at frequencies $\omega_1$ and $\omega_2$ are pinned around the thermal peak clusters corresponding to the two graphene drums G1 and G2 (Fig.~\ref{DTC_phases_quad}B). The centers of these drums (points $X_1$ and $X_2$) are separated by 40 micrometers. Therefore, by comparing the correlation between the peaks at $\omega_1$ and $\omega_2$, we can gauge the extent of spatial correlations in the system. For a single mode corresponding to either G1 or G2, quadrature fluctuations, measured in frame co-evolving at either of the two peak frequencies, show uncorrelated noise reminiscent of thermal equilibrium (Fig.~\ref{DTC_phases_quad}D, E).  In striking contrast, cross-correlation measurements between the quadrature of two different modes (effectively probing inter-drum correlations) show strong correlation, with a $24$ dB suppression along a specific phase compared to fluctuations in the orthogonal quadrature (Fig.~\ref{DTC_phases_quad}F, G), providing additional evidence of emergent long-range order in higher-order correlations or fluctuations. As expected, this correlation in fluctuations is absent below the DTC threshold (see SI, section I) and thereby, can be ascribed to it's emergent long-range order.

	%\section{Rigidity of DTC phases}

\begin{figure}
		\centering
		\includegraphics[scale=0.22]{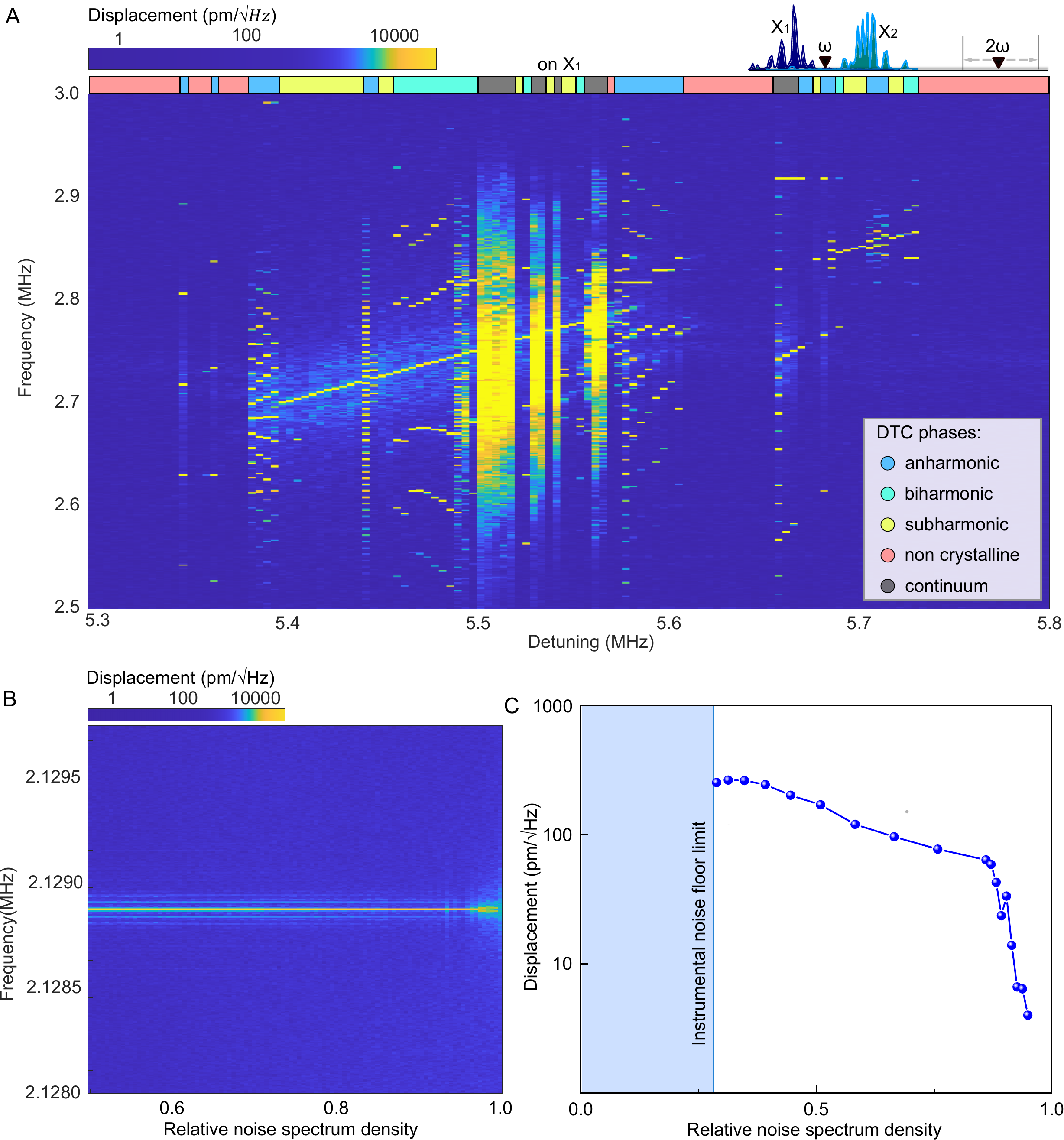}
  \caption{\textbf{Rigidity of DTC phases:} 
  \textbf{(A)} Experimental observation of displacement spectra at constant DC gate voltage vs parameter pump detuning. Pump frequency is detuned around twice of the frequency $\omega$ of the center of the clusters of hybrid modes (schematic in the upper right corner).  The emergent phases characterized by different symmetries are marked above. \textbf{(B)} Displacement spectrum density (along Y axis) of subharmonic peak is plotted against increasing noise amplitude (along X axis). \textbf{(C)} Subharmonic peak amplitude of Fig.~3B is plotted against increasing relative noise spectrum density. } 

  % \textbf{(B)} Cross section along Y axis of Fig.~3A color map is plotted for minimum (blue) and maximum (red) noise strength. 
		\label{DTC_rigidity}
	\end{figure}

  \paragraph*{\textbf{Rigidity against parameter variation:}}
Due to interplay between periodic drive, nonlinearity, linear and nonlinear damping of the system, the spectrum of the DTC phases remains stable for millions of oscillation periods over time scales of seconds. To test rigidity of the many-body system against parameter variation, we tune the drive frequency (at constant tension and drive power) and record  the resulting DTC spectra (Fig.~\ref{DTC_rigidity}A). Broad bands of DTC phases are observed with sharp transitions between them. In a remarkably rich phase diagram, we observe additional phases, distinguished by their symmetry, in addition to previously seen subharmonic, anharmonic and biharmonic phases. The continuum phase, characterized by a broadband spectrum, requires further investigation and can be a melted, chaotic or a coherent phase. The non-crystalline phase appears for certain pump detuning values which indicates that the drive strength is below threshold. This reflects the dependency of parametric threshold value on pump detuning and coupling of modes. Overall the phase diagram of our hybrid NEMS resonators with transitions between distinct DTC phases rivals the complexity of crystallographic phases of most solid state system.

 %    \paragraph*{\textbf{\hl{Rigidity against parameter fluctuation:}}} 
 % \hl{Due to interplay between periodic drive, nonlinearity, linear and nonlinear damping of the system, the spectrum of the DTC phases remains stable for millions of oscillation periods over time scales of seconds. To test rigidity of the many-body system against parameter fluctuation, we tune the drive frequency (at constant tension and drive power) and record  the resulting DTC spectra} (Fig.~\ref{DTC_rigidity}A). \hl{Broad bands of DTC phases are observed with sharp transitions between them. In a remarkably rich phase diagram, we observe additional phases, distinguished by their symmetry, in addition to previously seen subharmonic, anharmonic and biharmonic phases. The continuum phase, characterized by a broadband spectrum, requires further investigation and can be a melted, chaotic or a coherent phase. The non-crystalline phase appears for certain pump detuning values which indicates that the drive strength is below threshold. This reflects the dependency of parametric threshold value on pump detuning and coupling of modes. Overall the phase diagram of our hybrid NEMS resonators with transitions between distinct DTC phases rivals the complexity of crystallographic phases of most solid state system.}

     \paragraph*{\textbf{Rigidity against external noise:}} To test the rigidity of the many-body system against fluctuations, we mix a white noise signal into the drive channel and record the resulting DTC spectra with increasing noise amplitude (Fig.~\ref{DTC_rigidity}B). Fig.~\ref{DTC_rigidity}C shows a plot of the peak of the subharmonic spectra that remains stable till the relative noise becomes comparable to the drive and eventually \textit{melts} to a broad noisy spectrum. Remarkably, this transition is marked by a \textit{sharp} phase boundary at specific noise threshold, reminiscent of a first-order thermodynamic phase transition.

% (Fig.~\ref{DTC_rigidity}C). 
%In contrast to sensitivity of conventional phase synchronization to external parameters~\cite{Araki1975-bi, RevModPhys.77.137}, the observed DTC phases are stable with respect to parameter fluctuations. The rigidity to externally added noise is beyond predictions of effective two-mode mean-field models~\cite{PhysRevLett.117.214101,PhysRevLett.123.083901}. 
 
%\section{A mean-field model}
\paragraph*{\textbf{A mean-field model:}}
 Mean-field models of few coupled modes have been successful in providing insights into dynamical features of DTCs~\cite{PhysRevLett.123.124301,PhysRevLett.123.083901}. To investigate the nature of the observed phases further, we thereby model the system as two coupled nonlinear graphene modes (with transverse displacement $x_1$ and $x_2$) with a common parametric drive and open to a thermal bath with damping and fluctuations. The coarse graining model replaces the zoo of many coupled SiNx modes with a direct coupling between the two otherwise decoupled graphene drums. In effect, the model is represented by two damped Mathieu-Van der Pol-Duffing oscillators with bi-linear coupling between them (see SI, section II. A and ~\cite{Singh2019Giant}). Numerical simulations (see SI, section II. B) indicate the same three DTC phases emerging in different parts of the phase diagram (see appendix, Fig.~\ref{numerical_spectrum} ).

	\begin{figure}
		\centering
		\includegraphics[scale=0.13]{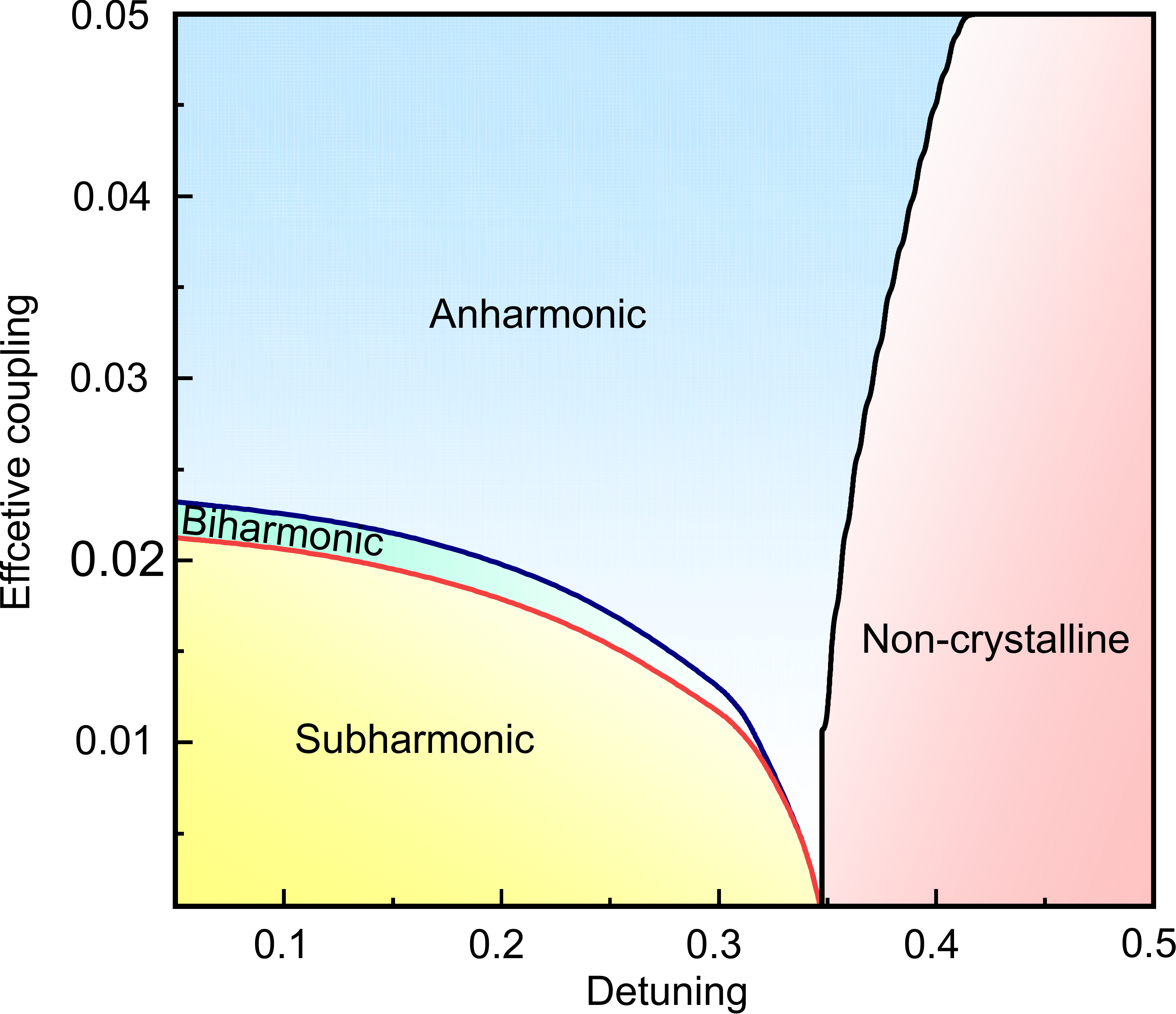}
		\caption{\textbf{Phase diagram of DTCs in the mean-field approximation:} Fixed point analysis in parameter space shows different possible phases of the mean-field model. Red, blue, green and yellow regions are representing non-crystalline, anharmonic, biharmonic and subharmonic responses respectively.}
        %Numerical spectrum of the \textbf{(A)} anharmonic phase,  \textbf{(B)} biharmonic phase and \textbf{(C)} subharmonic phase like responses (similar to Figs.~2A, B and C respectively) for specific values of parameters, coupling ($\alpha$) , detuning ($\delta$) and drive strength ($h$).

        \label{phase_diagram}
	\end{figure}
    
	This hints that few body classical models at least capture spectral features of genuine many-body DTC phases. A linear stability analysis of the model provides insights into the dynamical nature of the DTC phases.	Collective oscillatory DTC phase are mathematically represented as solutions of sinusoidal with frequency $\omega$. Amplitudes of the solutions are slowly varying owing to the weak nonlinearity assumed to facilitate the corresponding rigorous perturbative analyses (see SI, section II. C). The phases then correspond to robust stable attracting solutions observed for a set of initial conditions of finite measure, while difference in DTC phases manifest as different asymptotic forms of the slowly varying amplitudes. For high enough detuning, stable fixed point, indicating non-crystalline phase (red region, Fig.~\ref{phase_diagram}) appears at the origin (see appendix, Fig.~\ref{fixed_points}A, B). Reduction in detuning renders this stable solution unstable, eventually yielding limit cycle solutions symmetric about the origin (appendix, Fig.~\ref{fixed_points}E, F) via the Hopf bifurcation (black curve, Fig.~\ref{phase_diagram}). The corresponding steady-state oscillations have frequency shifted from $\omega$ resulting in anharmonic DTC phases (blue region, Fig.~\ref{phase_diagram})~\cite{PhysRevResearch.3.023106}. Remarkably, as the coupling strength is further decreased, the limit cycle too loses its stability. However, it bifurcates to give rise to two new stable limit cycles that are not symmetric about origin (appendix, Fig.~\ref{fixed_points}G), leading to a novel unexplored scenario where DTC spectra bears both the spectral features at frequency $\omega$ as well as the frequencies symmetrically shifted from $\omega$. This suggests a new DTC phase: bi-harmonic (green region, Fig.~\ref{phase_diagram}). The green region indicates existence of more complex solutions that need further detailed analysis. As the coupling strength is further reduced, the two limit cycles eventually disappear leaving only stable fixed points with constant amplitude, shifted away from the origin (appendix, Fig.~\ref{fixed_points}C, D). This case corresponds to subharmonic entrainment (yellow region, Fig.~\ref{phase_diagram}) with frequency at $\omega$ ~\cite{Zhang2017,Choi2017,Autti2022,doi:10.1126/sciadv.aay8892}. Higher order effects of nonlinearities further complicate the spectra through a series side-peaks in the frequency spectra. \\
    Although, the mean-field model provides crucial insights into the dynamical nature of the DTC phases, and dependence to a coarse-grained coupling and mode detuning, the rigidity to externally added noise is beyond predictions of effective two-mode mean-field models~\cite{PhysRevLett.117.214101,PhysRevLett.123.083901} (see appendix, Fig.~\ref{numerical_rigidity})).

	%\section{Outlook}
\paragraph*{\textbf{Outlook:}}	
	Our results establish existences of truly many-body DTCs in classical systems with distinctly different phases. Observation of rigidity against parameter variations and sharp phase boundaries are found to be emergent many-body features beyond mean-field paradigm. Furthermore, the hybrid NEMS, with exceptional nonlinear and elastic properties of graphene resonators are found to be essential in providing tunable interactions, nonlinearity and damping between many SiNx modes resulting in the stability of the DTC phases. 
	
	\par These results imply multiple directions to explore. Unexplained spectral features (Region II of Fig.~\ref{device}D and continuum phase of Fig.~\ref{DTC_rigidity}A), can lead to understanding of liquid-like phases in time~\cite{PhysRevLett.127.140602,doi:10.1126/science.abg2530,doi:10.1126/science.abg8102} and the nature of eventual thermalization in these non-equilibrium systems~\cite{PhysRevLett.127.140602,PhysRevLett.127.140603}. Correspondence of these DTC phases with their quantum counterparts~\cite{PhysRevLett.127.140603,PhysRevLett.127.140602} can lead to insights into interacting quantum systems. For NEMS, multiple graphene resonators in spatially ordered geometries can be extended to new space-time-crystalline phases, exceptional spectral rigidity of DTCs can be applied as unique frequency markers in information technology~\cite{https://doi.org/10.48550/arxiv.2206.11661,Faust2013,Sillanpaa} while tunability to phase boundaries can be used in sensors and metrology~\cite{PhysRevLett.128.080603, PhysRevE.94.022201,Weber2016,Dolleman2016,https://doi.org/10.48550/arxiv.2205.11903}.

%\section{Acknowledgments}	  
\paragraph*{\textbf{Acknowledgments:}}
We thank Adhip Agarwala, Souvik Bandyopadhyay, Sayak Bhattacharjee, Diptarka Das, Amit Dutta, Shilpi Gupta and V. Prasanna for illuminating discussions.
This work is supported by SERB, Department of Science and Technology, India under DST grant no: SERB/PHY/2015404, ERC Grant No. 639739 and DFG, German Research Foundation, projects 449506295 and 328545488, CRC/TRR 227. A.S acknowledges CSIR, New Delhi and A.K.R acknowledges MHRD for financial support. J.A.M. acknowledges Prime Minister's Research Fellows (PMRF) scheme of the Ministry of Human Resource Development, Govt. of India for financial support.

 \newpage
% \onecolumngrid\newpage\twocolumngrid

        \section*{APPENDIX: Features of mean-field model}
         % \paragraph*{\textbf{Features of mean-field model:}}\\
        % \paragraph*{\textbf{Different DTC like spectra of mean-field model:}}

%\section{Fixed point analysis of mean-field model}
  \textbf{1. Different DTC like spectra:}
\begin{figure}[H]
	\centering
	\includegraphics[scale=0.12]{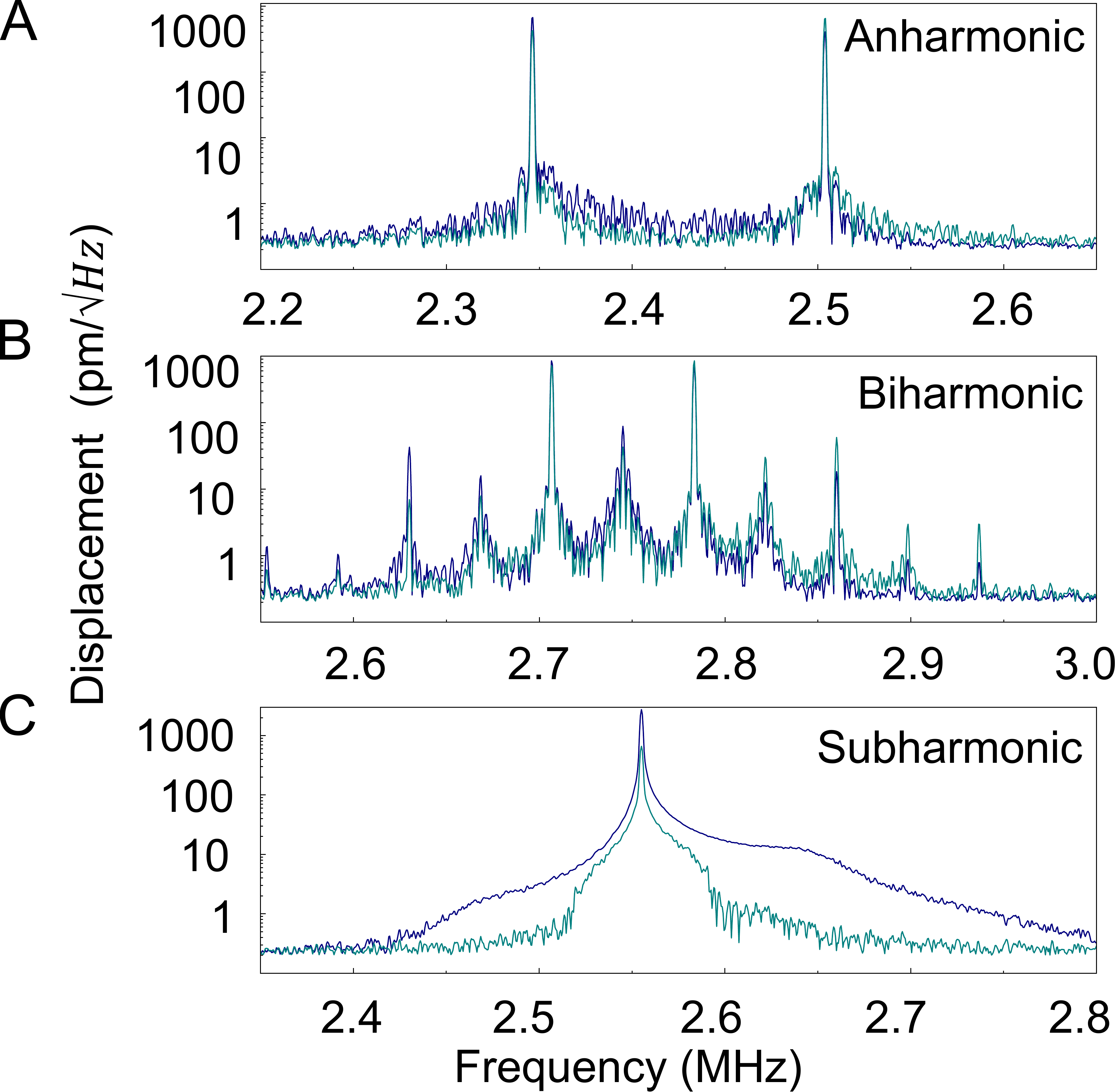}
	\caption{\textbf{Different DTC phases :} Numerical spectrum of the \textbf{(A)} anharmonic phase,  \textbf{(B)} biharmonic phase and \textbf{(C)} subharmonic phase like responses (similar to Figs.~2A, B and C respectively) for specific parameter values.} 
	\label{numerical_spectrum}
\end{figure}

\begin{figure*}
	\centering
	\includegraphics[scale=0.2]{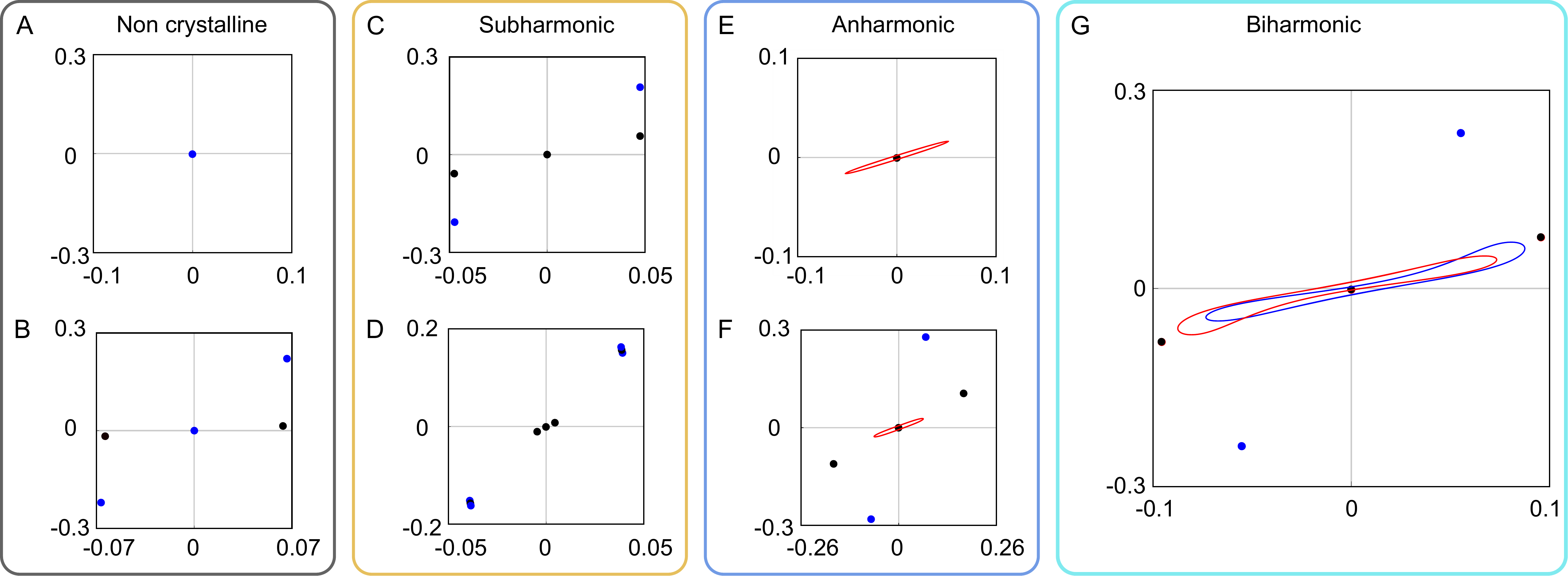}
	\caption{\textbf{Fixed point analysis:} Non crystalline phase appears when \textbf{(A)} system has one (blue dot) stable fixed point at $(0,0)$ or \textbf{(B)} five fixed points, among which $(0,0)$ is a stable fixed point. Two other stable fixed points are far away from origin. Other two black fixed points are unstable. \textbf{(C)} Subharmonic phase is observed in presence of five fixed points among which $(0,0)$ is an unstable fixed point. Two stable fixed points are far away from origin. System reaches to one of the fixed points when dynamic stability is reached. Other two black dots are also unstable. \textbf{(D)} Five of nine fixed points including $(0,0)$ are unstable. Four stable fixed points are away from origin. System reaches to one of the fixed points when dynamic stability is reached.  \textbf{(E)} Anharmonic phase is observed when origin is unstable fixed point and surrounded by a stable limit cycle. \textbf{(F)} In presence of other stable and unstable fixed points outside limit cycle, system shows anharmonic phase. \textbf{(G)} When limit cycle becomes asymmetric, and two possible limit cycle can appear depending on initial conditions, system shows biharmonic response.} 
	\label{fixed_points}
\end{figure*}

 % \paragraph*{\textbf{Fixed point analysis:}}
 \textbf{2. Fixed point analysis:} Using the multiple timescales analysis, we find the first order differential equations for real and imaginary parts of amplitudes of our mean-field model (see SI, section II. C). We are now all set to study the dynamics of the system in four dimensional phase space via linear stability analysis. In Fig.~\ref{fixed_points}, we show various dynamical phases of the system in two dimensional planes at different sets of parameter values, where blue and black dots are stable and
unstable fixed points respectively. Elliptical paths around an unstable fixed point represent limit cycles.  

\vspace{5mm} 
\textbf{Non crystalline phase:} In Fig.~\ref{fixed_points}A and Fig.~\ref{fixed_points}B, we show the fixed points corresponding to the non crystalline phase of the system. In both the cases, origin in is the stable fixed point. As in Fig.~\ref{fixed_points}A, there exists only one fixed point at origin which is stable, therefore any initial condition will end up at this point and there will be no amplification and it corresponds to the non crystalline phase. In Fig.~\ref{fixed_points}B also, origin is stable fixed point and the other fixed point are far away from it therefore in small amplitude regime it corresponds to the  noncrystalline phase. 

\vspace{5mm} 
\textbf{Subharmonic:} In Fig.~\ref{fixed_points}C and Fig.~\ref{fixed_points}D, we depict the fixed points corresponding to the subharmonic phase of the system. In this case, origin becomes unstable fixed point along with other stable fixed points. Fixed point diagram for subharmonic phase in Fig.~\ref{fixed_points}C have four other fixed points (other than zero) in which two (in blue) correspond to the stable and two (in black) correspond to the unstable one. However, the subharmonic phase in Fig.~\ref{fixed_points}D have eight other fixed points (other than zero) in which four (in blue) corresponds to the stable and four (in black) corresponds to the unstable one.

\vspace{5mm} 
\textbf{Anharmonic:} Fig. \ref{fixed_points}E and Fig.~\ref{fixed_points}F depict the fixed points corresponding to the anharmonic phase of the system. In the anharmonic phase, the unstable origin is enclosed by a limit cycle. Limit cycle is symmetrically placed around the origin. Fixed point diagram for anharmonic phase in Fig. \ref{fixed_points}E have only one unstable fixed point (origin) and a limit cycle enclosing it. Whereas fixed point diagram for anharmonic phase in Fig. \ref{fixed_points}F not only have one unstable fixed point (origin) and a limit cycle enclosing it but also there are four other fixed points in which two (in black) are unstable and other two (in blue) are stable.

\vspace{5mm} 
\textbf{Biharmonic:} Fig. \ref{fixed_points}G represents the fixed points diagram corresponding to the biharmonic phase of the system. In the biharmonic phase, the unstable origin is enclosed by two limit cycles  asymmetrically placed around the origin.

The biharmonic phase can be characterized in two ways. Firstly, in the biharmonic phase one finds the appearance of a peak in the spectrum at half of the drive frequency $\omega$ (Fig. \ref{numerical_spectrum}B). Secondly, in the quadrature space, as we tune the parameters to move from the anharmonic to biharmonic phase(Fig. \ref{numerical_spectrum}B), one finds that the corresponding limit cycle breaks the reflection symmetry about the origin. Consequently, there are two limit cycle solutions to the perturbative equation connected by the parity transformation ($(a_r, a_i, b_r, b_i)\xrightarrow{}(-a_r, -a_i, -b_r, -b_i)$) of the four dimensional phase space (See Fig. \ref{fixed_points}G). This also has the implication that mean position of the limit cycle is no longer at the origin of the quadrature space. In order to see the connection between two characterizations, consider one of the displacements, say $x(t)$. To keep the argument clear, we indulge in abuse of notation and ignore complex conjugation terms to write $x(t)=a(t)e^{i\omega t}$. Now, since for a limit cycle $a(t)$ is periodic, one may write it as a Fourier series,
\begin{equation}
	a(t)=a_0+a_1e^{i\omega_1t}+a_2e^{i\omega_2t}+\dots
\end{equation} 
Thus the displacement takes a form
\begin{equation}
	x(t)=a_0e^{i\omega t}+a_1e^{i(\omega_1+\omega)t}+a_2e^{i(\omega_2+\omega)t}+\dots
\end{equation} 
This equation says that if the quadrature space limit cycle contains a frequency component $\omega_i$, there is a peak in the spectrum of $x$ at $\omega+\omega_i$. As mentioned above, in the biharmonic phase limit cycle is not symmetric about the origin and thus at least one component in the quadrature space has a nonzero time average. A nonzero time average in turn implies a zero frequency mode, or $a_0\neq0$. Thus $a$ has an $\omega_i=0$ mode which in turn means that $x$ has a peak at $\omega+0=\omega$.

We remark here that the breaking of the reflection or $\mathbb{Z}_2$ symmetry by the limit cycle suggests a connection to Landau's theory of ferromagnetism. Particularly, the distance of the limit cycle center from the origin functions effectively as an order parameter for the transition. Moreover, the direct forcing term that we have ignored in the analysis may function as the analogue of the conjugate field coupling to the order parameter. We leave this analogy here for study in a future work.
\\

\textbf{3. Rigidity against increasing fluctuation:}

\begin{figure}[H]
	\centering
	\includegraphics[scale=0.12]{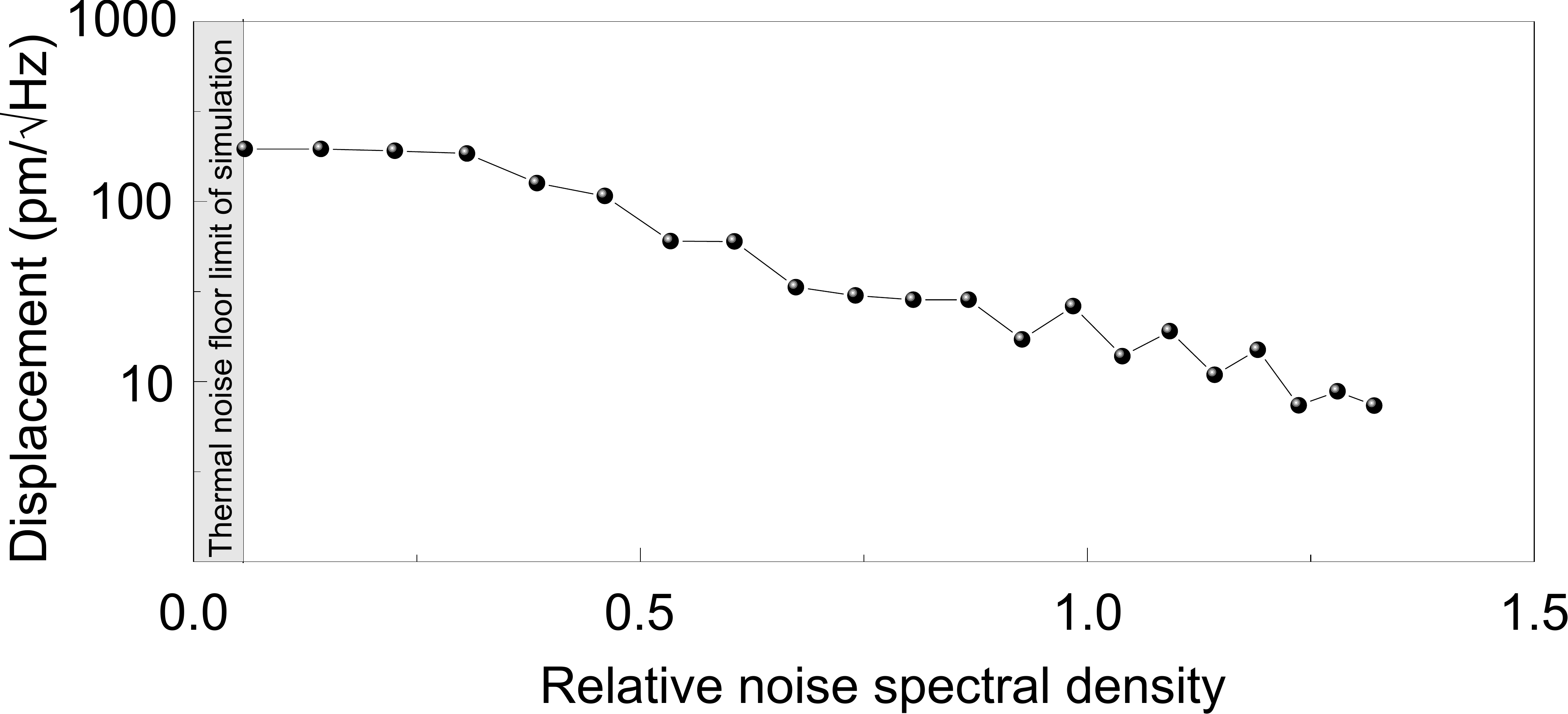}
	\caption{Numerically extracted subharmonic peak amplitude is plotted against increasing relative noise spectrum density.} 
	\label{numerical_rigidity}
\end{figure}

 In contrast to experimental result, numerical simulation of mean field model shows rapid fluctuation between the DTC and a melted phase above a certain noise strength. When averaged over many runs, no sharp transition is observed (Fig.~\ref{numerical_rigidity}). We interpret the stark difference between the experimental system and a  mean-field model as a direct evidence for the many-body nature of the emergent DTC phase.

 \newpage

\section{Evidence of correlated fluctuations and long-range order in anharmonic phase}

\begin{figure}[H]
	\centering
	\includegraphics[scale=0.25]{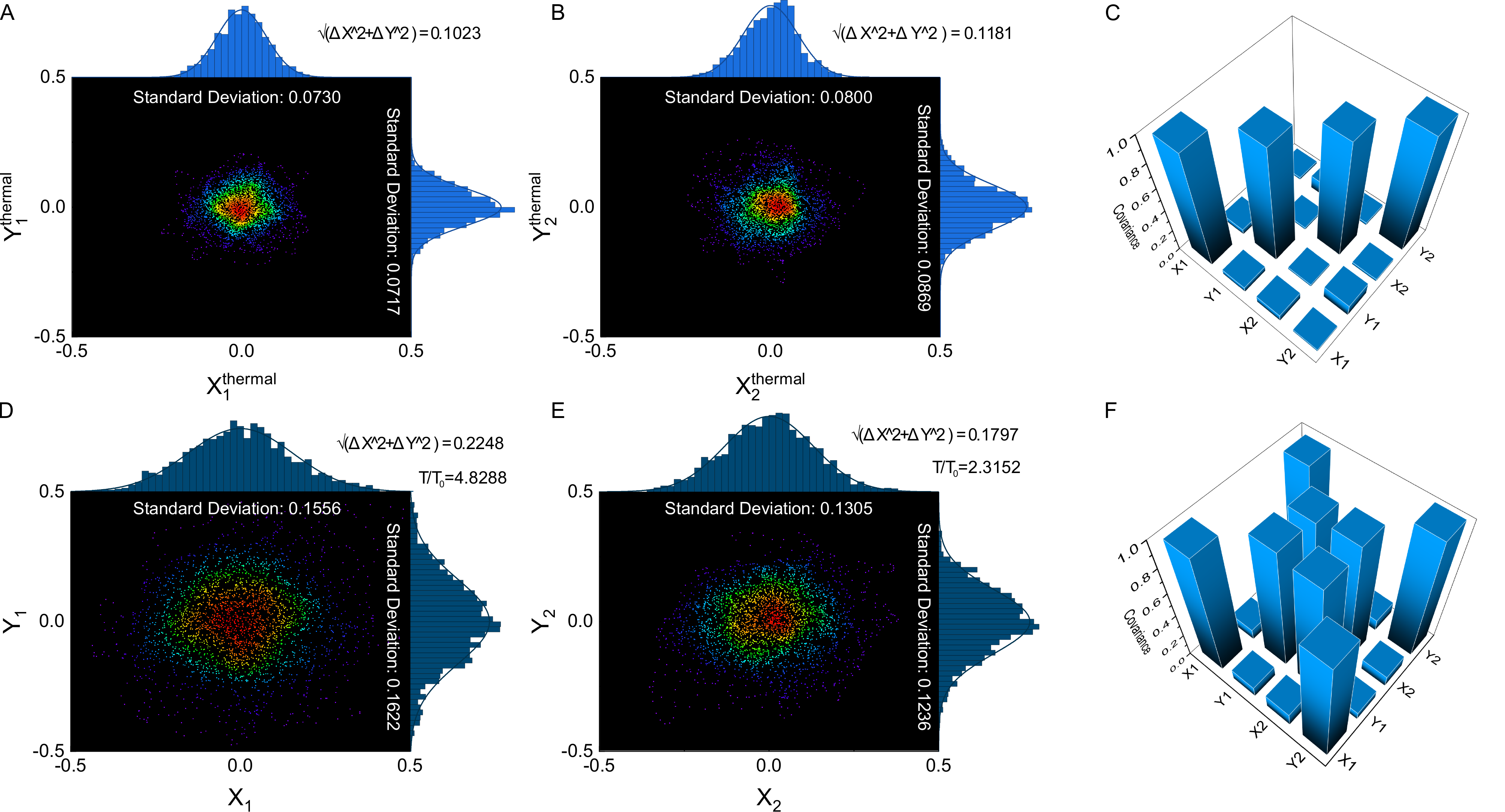}
	\caption{\textbf{Correlation:} Quadrature plots with histograms of \textbf{(A)} peak at $\omega_1$ and \textbf{(B)} peak at $\omega_2$ show noise distribution when system is not driven. \textbf{(C)} Plotting covariance matrix we observe no cross-correlation between fundamental mode frequencies when measured on $X_1$. \textbf{(D)},\textbf{(E)} and \textbf{(F)} are similar to \textbf{(A)},\textbf{(B)} and \textbf{(C)} respectively when system is driven and showing anharmonic response.} 
	\label{quad_temp}
\end{figure}

To measure long-range correlation between two spatially separated measurement points on the surface, we focus on the anharmonic phase in details, where two distinct peaks appears at $\omega_1$ and $\omega_2$ due to drum 1 ($X_1$) and drum 2 ($X_2$) respectively. Quadrature measurement using a network analyzer is done to detect fluctuations at $\omega_1$ and $\omega_2$.  We define in phase and out of phase components of peak $\omega_1$ and $\omega_2$ as $X_1,Y_1,X_2$ and $Y_2$, respectively. In the absence of parametric pumping, each mode shows thermal noise distribution (see Fig. \ref{quad_temp}A and B ) in phase space due to room temperature.

Furthermore, to quantify the correlation between modes, we calculate the absolute correlation coefficients for {$X_1,Y_1,X_2,Y_2$}, defined as \cite{PhysRevLett.113.167203}

\begin{equation}
	C_{i,j}(\tau)=|cov(Z_i,Z_j(\tau))/\sigma_{Z_i}\sigma_{Z_j}(\tau)|.
\end{equation}

Here, $cov(Z_i,Z_j(\tau))= \sum_{k}(Z_{i,k}-\bar{Z_i})(Z_{j,k}(\tau)-\bar{Z_j})/N$, $N$ is total number of data points, $Z_i\in \{X_1,Y_1,X_2, Y_2\}$ , $\sigma_{Z_i}$ is the variance and $\tau$ is time delay between two data points.

Fig. \ref{quad_temp}C is showing absolute correlation coefficients at $\tau=0$ correspond to thermal peak. When the system is not driven, We observe zero cross-correlation terms, indicating absence of correlation between quadrature fluctuations of spatially separated measurement points. However, under parametric pumping, overall fluctuation increases and cross quadrature shows anti-correlation (Fig. 2F and G in main text). This reveals strong cross-mode correlations, reflected in non-zero off-diagonal elements of covariance matrix (Fig. \ref{quad_temp}F) .  

\begin{figure}
	\centering
	\includegraphics[scale=0.38]{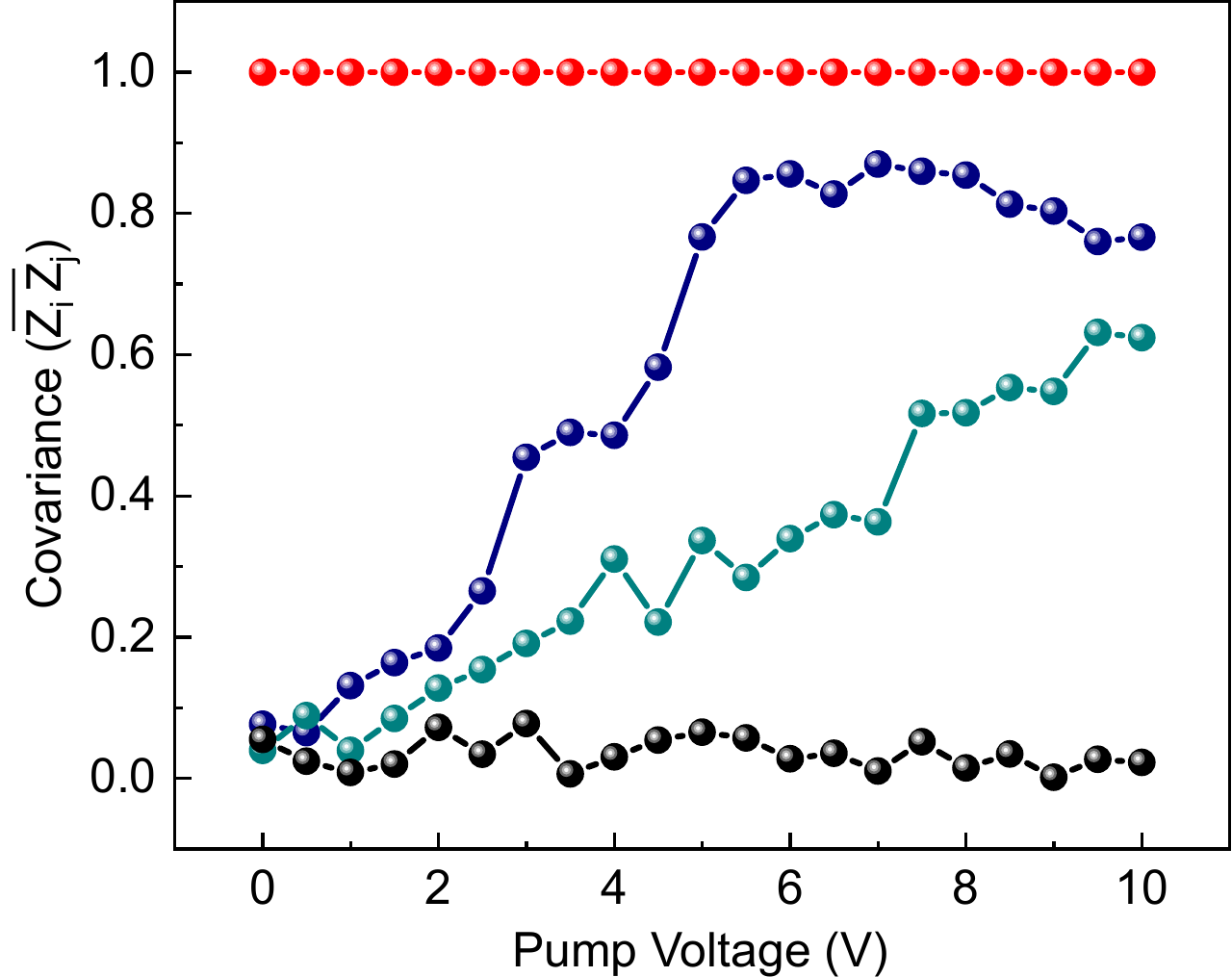}
	\caption{\textbf{Variation of covariance with increasing pump amplitude}}
	\label{quad_pump} 
\end{figure}

Further, as pump amplitude increases, covariance $X_1-X_2$ (navy blue) and $X_1-Y_2$ (dark cyan) increase linearly (Fig. \ref{quad_pump}), but with different slope. Navy blue curve saturates around $0.8$ and starts decreasing again. On the other hand, covariance $X_1-Y_1$ (black) and $X_1-X_1$ (red) remain zero and one respectively for all the time with increasing pump strength.

\newpage
\section{Details of the mean-field model}
\subsection{Effective Hamiltonian of the system}

Fig. 1B in main text shows that the fundamental mode of graphene drums can have resonance coupling with multiple high quality factor, densely packed higher order $SiNx$ modes. While both of the graphene drums couple to a few substrate modes on resonance, there is no direct graphene-to-graphene mode coupling
% (no signature of avoided level-crossing in Fig. 1C in main text).
. Similarly, coupling among $SiNx$ modes exist through graphene-mode mediated interaction.

Hamiltonian of this coupled system can be represented as

\begin{equation}
	\mathcal{H}= \frac{p_1^2}{2m_g}+ \frac{p_2^2}{2m_g}+\sum_{i=3}^{n}\frac{p_i^2}{2m_s}+\frac{1}{2}m_g \omega_1^2 x_1^2+ \frac{1}{2}m_g \omega_2^2 x_2^2 + \frac{1}{2}m_s \sum_{i=3}^{n} \omega_i^2 x_i^2+\frac{1}{4} \beta (x_1^4+ x_2^4)- \sum_{i=3}^{n} \alpha'_{1i}x_1x_i-\sum_{i=3}^{n}\alpha'_{2i}x_2 x_i.
\end{equation}	

%	For simplicity, let's assume that there is only one $SiNx$ mode, which is interacting with fundamental modes of both the graphene drums.

We can write following differential equations for this system:

\begin{subequations} \label{equation 8}
	\begin{eqnarray}
		&&\ddot{x}_1+\omega_1^2 x_1+\frac{\beta}{m_g}  x_1^3-\sum_{i=3}^{n}\frac{ \alpha'_{1i}}{m_g} x_i=0\,,\\
		&&\ddot{x}_2+\omega_2^2 x_2+\frac{\beta}{m_g}  x_2^3-\sum_{i=3}^{n}\frac{ \alpha'_{2i}}{m_g} x_i=0\,,\\
		&&\ddot{x}_i+\omega_i^2 x_i-\frac{\alpha'_{1i}}{m_s} x_1-\frac{\alpha'_{2i}}{m_s}x_2=0\,, \quad\forall i=\{3,4,\cdots,n\}.
	\end{eqnarray}
\end{subequations}
Now, we know that $SiNx$ is much heavier than graphene membrane $(m_s\gg m_g)$. So displacement of graphene drums are order of magnitude larger than $SiNx$. Observing thermal peaks at fundamental frequencies we take ansatz $x_1\sim\cos(\omega_1 t)$ and $x_2\sim\cos(\omega_2 t)$.  In this situation, we will focus on $SiNx$ equations. Because secular terms decay down very fast, only coupling terms act as driving forces for this equation, implying $x_i\sim A_i\cos(\omega_1 t)+B_i\cos(\omega_2 t)\sim A'_ix_1+B'_ix_2$ ($A_i$, $B_i$, $A'_i$, and $B'_i$ are some slow variables of time) $\quad\forall i=\{3,4,\cdots,n\}$.

If we plug in this into equation \ref{equation 8} a and b, we can rewrite equations,
\begin{subequations}\label{equation 9}
	\begin{eqnarray}
		\label{nl_HO}
		&&\ddot{x}_1 +\omega_1^2 x_1 + \frac{\beta}{m_g}  x_1^3 - \frac{\alpha}{m_g} x_2=0\,,\\
		&&\ddot{x}_2 +\omega_2^2 x_2 + \frac{\beta}{m_g}  x_2^3 - \frac{\alpha}{m_g} x_1=0\,.
	\end{eqnarray}
\end{subequations}
One of the two coupling terms will get absorbed as fundamental frequency correction terms, and finally we can think of an effective coupling between two graphene drums.

We write effective Hamiltonian for the system,

\begin{equation}
	\mathcal{H}_{eff}=  \frac{p_1^2}{2m_g}+ \frac{p_2^2}{2m_g}+ \frac{1}{2}m_g \omega_1^2 x_1^2+\frac{1}{2} m_g \omega_2^2 x_2^2 + \frac{1}{4} \beta(x_1^4+ x_2^4)- \alpha x_1 x_2.
\end{equation}

In our experiments, we take $\omega_1=\omega_0-\Delta/2$ and $\omega_2=\omega_0+\Delta/2$, and parametrically drive the system with frequency $2\omega_0+\Omega_p$ such that the Hamiltonian becomes:
\begin{eqnarray}
	&&	{H}=  \frac{p_1^2}{2m_g}+ \frac{p_2^2}{2m_g}+ \frac{1}{2}m_g \left(\omega_0-\frac{\Delta}{2}\right)^2 x_1^2+\frac{1}{2}h \cos(\left\lbrace2\omega_0+\Omega_p\right\rbrace t+\chi_1) x_1^2\\\nonumber
	&&\phantom{\mathcal{H}=}+\frac{1}{2}m_g \left(\omega_0+\frac{\Delta}{2}\right)^2 x_2^2+\frac{1}{2}h \cos(\left\lbrace2\omega_0+\Omega_p\right\rbrace t+\chi_2) x_2^2 +\beta(x_1^4+ x_2^4)- \alpha x_1 x_2.
\end{eqnarray}

%%%%%%%%%%%%%%%%%%%%%%%	
{\bf Estimation of coupling value for simulation:} Estimation of this effective coupling value will guide us to do numerical calculations. For $h=0$, we will try to solve Eq.~(\ref{equation 9}) by solving eigenvalue problem of the following matrix:
\begin{equation}
	\begin{bmatrix}
		\left(\omega_0-\frac{\Delta}{2}\right)^2 & -\frac{\alpha}{m_g} \\  -\frac{\alpha}{m_g} & \left(\omega_0 +\frac{\Delta}{2}\right)^2 
	\end{bmatrix}.
\end{equation}

Eigenvalues of this matrix are,
\begin{equation}
	{\omega^2}_\pm= \left({\omega_0}^2+{(\Delta /2)}^2\right)\left[1\pm\frac{\omega_0^2\Delta^2+(\alpha/m_g)}{\left({\omega_0}^2+{(\Delta /2)}^2\right)}\right],
\end{equation}
or,
\begin{equation}
	\delta \omega=\sqrt{\frac{\omega_0^2\Delta^2+(\alpha/m_g)}{{\omega_0}^2+{(\Delta /2)}^2}}, 
\end{equation}
where $\delta \omega=\omega_+ - \omega_-$.

Due to small coupling $\alpha$, we assume that the separation between the peaks at $\omega_+$ and $\omega_-$ is approximately equal to the detuning $\Delta$, we arrive at
\begin{equation}
	\alpha \simeq \frac{m_g \Delta^2}{2},
\end{equation}

for $m_g \approx{10}^{-15}$ and $\Delta \approx{10}^{5} $ 
we get $ \alpha \approx{10}^{-4}$.

Fine tuning of	$\alpha$ can be done by looking at experimental data.

\subsection{Numerical integration of stochastic differential equation}

As this system is in contact with a thermal bath at the room temperature, we can introduce effective damping and thermal noise $f$ and write coupled second order differential equations:

\begin{subequations} \label{mean_field}
	\begin{eqnarray}
		&&\ddot{x}_1+ \gamma \dot{x}_1 + \omega_1^2 x_1+  \frac{h}{m_g} \cos\left\{(\omega_1+\omega_2) t+\chi_1\right\} x_1
		+ \frac{\eta}{m_g}x_1^2 \dot{x}_1 +  \frac{\beta}{m_g} x_1^3-\frac{\alpha}{m_g} x_2=f_1(t),\\
		&&\ddot{x}_2+ \gamma \dot{x}_2 + \omega_2^2 x_2+ \frac{h}{m_g}  \cos\left\{(\omega_1+\omega_2) t+\chi_2\right\} x_2 + \frac{\eta}{m_g}x_2^2 \dot{x}_2 + \frac{\beta}{m_g} x_2^3-\frac{\alpha}{m_g} x_1=f_2(t).
	\end{eqnarray}
\end{subequations}

Conventional ODE solvers alone are inadequate to solve these coupled-oscillator equations due to the presence of stochastic terms. Moreover, an additional noise term is introduced while testing robustness of the simulated responses. As a result, Milstein algorithm is used for numerical integration to solve the equations numerically. 

%In cases of multiplicative noise, more accurate schemes such as the \textbf{Milstein algorithm} are used to account for the extra drift terms arising from the dependence of $g$ on $x$. However, because our thermal noise is effectively additive (see Eq.~(S.\,???) in the main text), we employ Euler--Maruyama in practice, which yields consistent results without requiring further corrections. For additional technical details, see Ref.~[\dots]

\subsubsection{Algorithm to solve stochastic differential equation (SDE)}
Let us take a first order differential equation with deterministic and stochastic term ~\cite{doi:https://doi.org/10.1002/9783527683147.ch7}:

\begin{equation}
	\frac{dx}{dt}=q(x,t)+ g(x,t) \zeta (t),
\end{equation}

where $q(x,t)$ and $g(x,t)$ are differentiable functions and  $\zeta (t)$ is Gaussian white noise ($<\zeta (t)>=0$ and \newline $<\zeta (t_1) \zeta (t_2)>= \delta(t_1-t_2)$). 

Following standard approach, we get the the recurrence relation known as the \textbf{Milshtein algorithm}:
\begin{eqnarray} \label{milstein}
	x(t+h)= x(t)+ h q(x(t),t) + g(x(t),t) \sqrt{h} u(t) + \frac{h}{2} g(x(t),t) g'(x(t),t) u^2 (t).
\end{eqnarray}
For purely additive noise ( $g$ is constant), this simplifies to the \textbf{Euler--Maruyama alogirthm} :
\begin{equation}
	x(t+h) = x(t) + h\,q\bigl(x(t),t\bigr) + D\,\sqrt{h}\,u(t),
\end{equation}

Here D is a constant represents additive noise strength.

In equation \ref{mean_field}, thermal fluctuation term is additive, so Euler-Maruyama algorithm is sufficient for our numerical integration.

\subsubsection{Numerical results for DTC phases} 

Solving equation \ref{mean_field} using the described stochastic algorithm, we obtain different spectra for different set of parameter values that resembles the experimentally observed subharmonic, anharmonic, and biharmonic DTC phases. While these results capture the essential features of the time crystalline phases, it fails to capture certain many-body aspects of our system (Fig. 7 in main text).

\begin{figure}[H]
	\centering
	\includegraphics[scale=0.29]{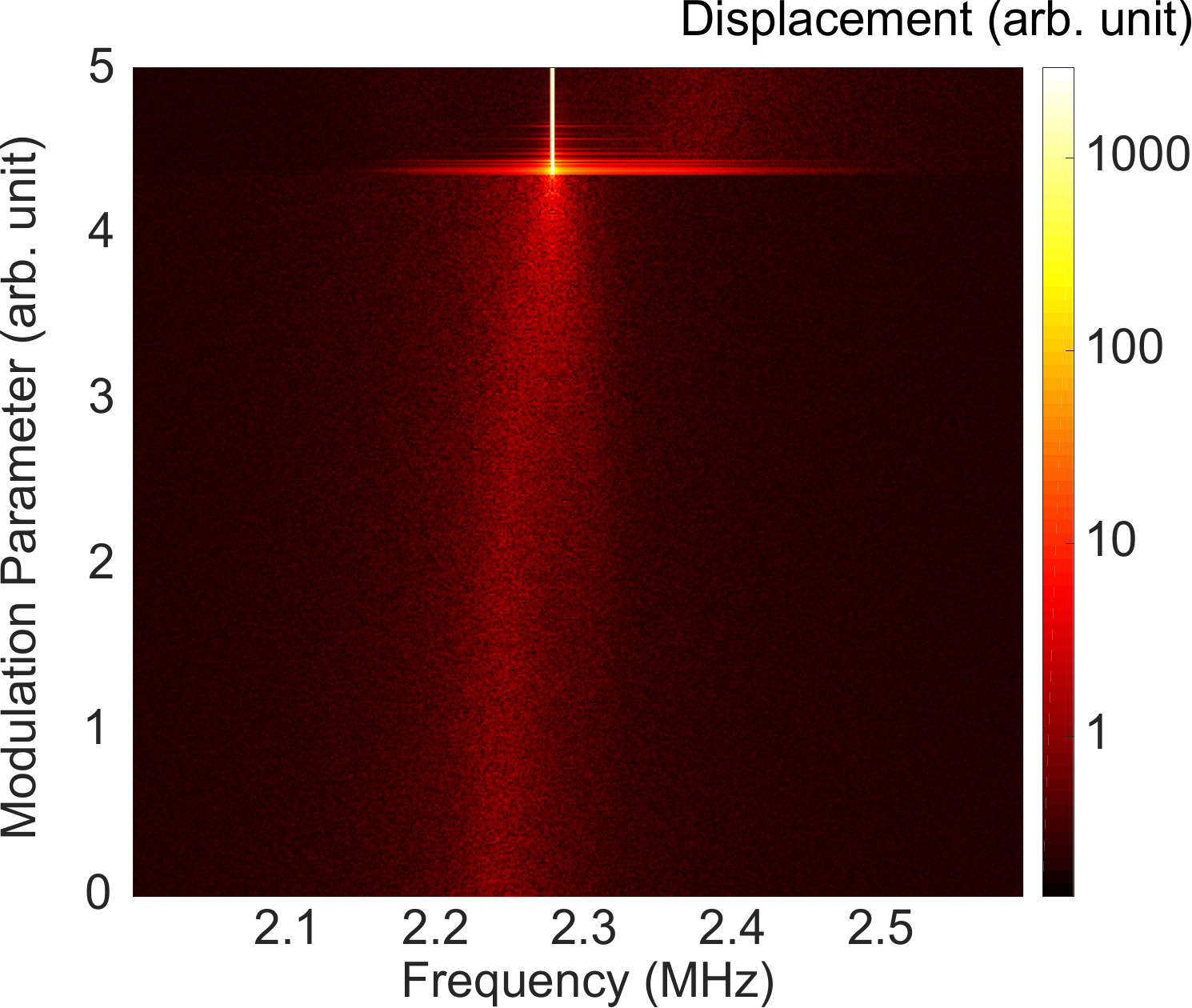}
	\caption{\textbf{Subharmonic phase:} Simulated spectrum shows appearance of subharmonic phase with increasing modulation parameter and keeping other parameters fixed.}
	\label{subharmonic_2d} 
\end{figure}

\vspace{5mm} 
\textbf{Subharmonic:} In experiment, at higher DC gate voltage fundamental modes of graphene drums are close enough to show subharmonic phase when driven parametrically. All the contributing modes start oscillating at half of the pump frequency (Fig.~2A, main text). Above a threshold value ($8 V$) of pump voltage we observe subharmonic response.  Using equations \ref{mean_field} with suitable parameters we simulate similar behavior (see Fig. \ref{subharmonic_2d}).

\begin{figure}[H]
	\centering
	\includegraphics[scale=0.3]{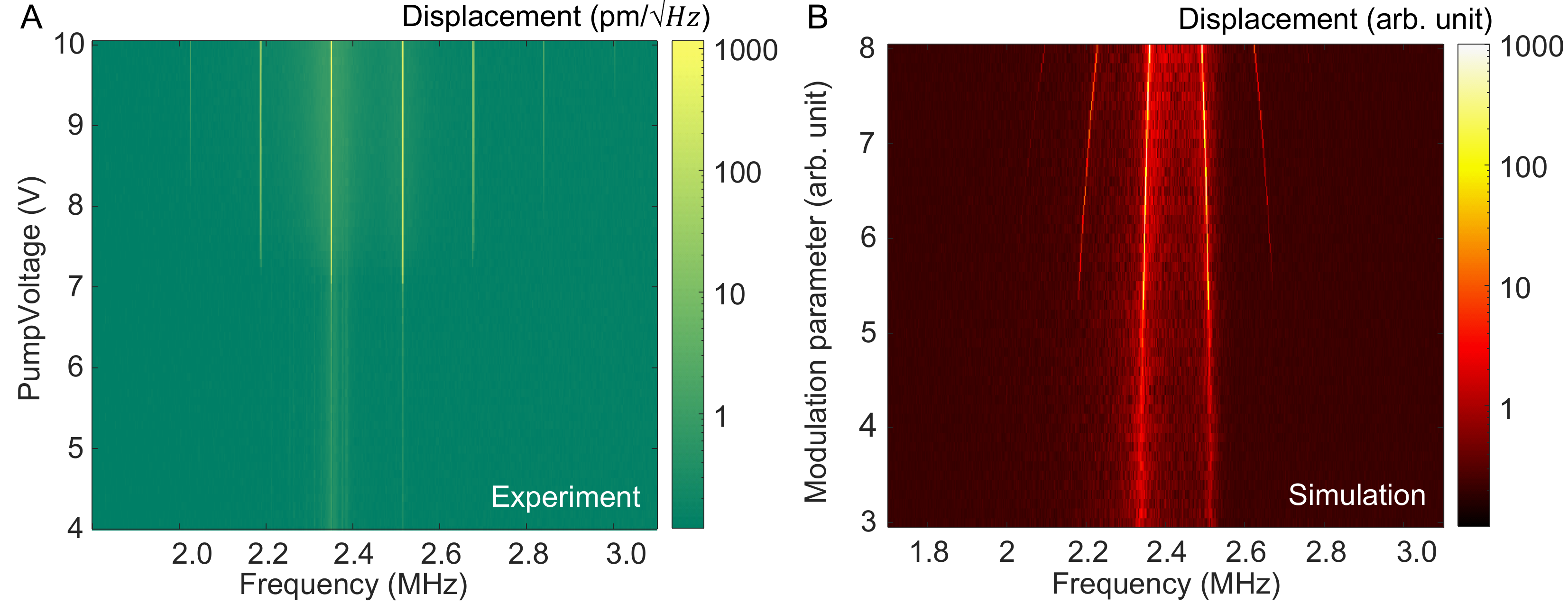}
	\caption{\textbf{Anharmonic phase:} \textbf{(A)} Anharmonic phase on $X_1$ with increasing pump voltage. \textbf{(B)} Numerical result with same conditions on $X_1$.} 
	\label{anharmonic_2d}
\end{figure}

\vspace{5mm} 
\textbf{Anharmonic:} We observe transition from noncrystalline phase to anharmonic phase by changing pump voltage when system is driven at $\omega_1+\omega_2$. Frequency spectrum (see Fig. \ref{anharmonic_2d}A) on $X_1$ is recorded. We simulate similar result (Fig. \ref{anharmonic_2d}B) using coupled nonlinear equations (\ref{mean_field}). 

%\begin{subequations}
%	\begin{eqnarray}
	%		\label{eq_13a}
	%		&&\ddot{x}= -\gamma\dot{x}-\frac{\eta}{m_g}x^2 \dot{x}-\left((\omega-\frac{\Delta}{2})^2+\frac{h}{m_g} \cos\left((2\omega+\omega_p)t+\chi_x\right)\right)x-\frac{\beta}{m_g} x^3+\frac{\alpha}{m_g} y	\\
	%		\label{eq_13b}
	%		&&\ddot{y}= -\gamma\dot{y}-\frac{\eta}{m_g}y^2 \dot{y}-\left((\omega+\frac{\Delta}{2})^2+\frac{h}{m_g} \cos((2\omega+\omega_p)t+\chi_y)\right)y-\frac{\beta}{m_g} y^3+\frac{\alpha}{m_g} x	
	%	\end{eqnarray}
%\end{subequations}

\begin{figure}[H]
	\centering
	\includegraphics[scale=0.7]{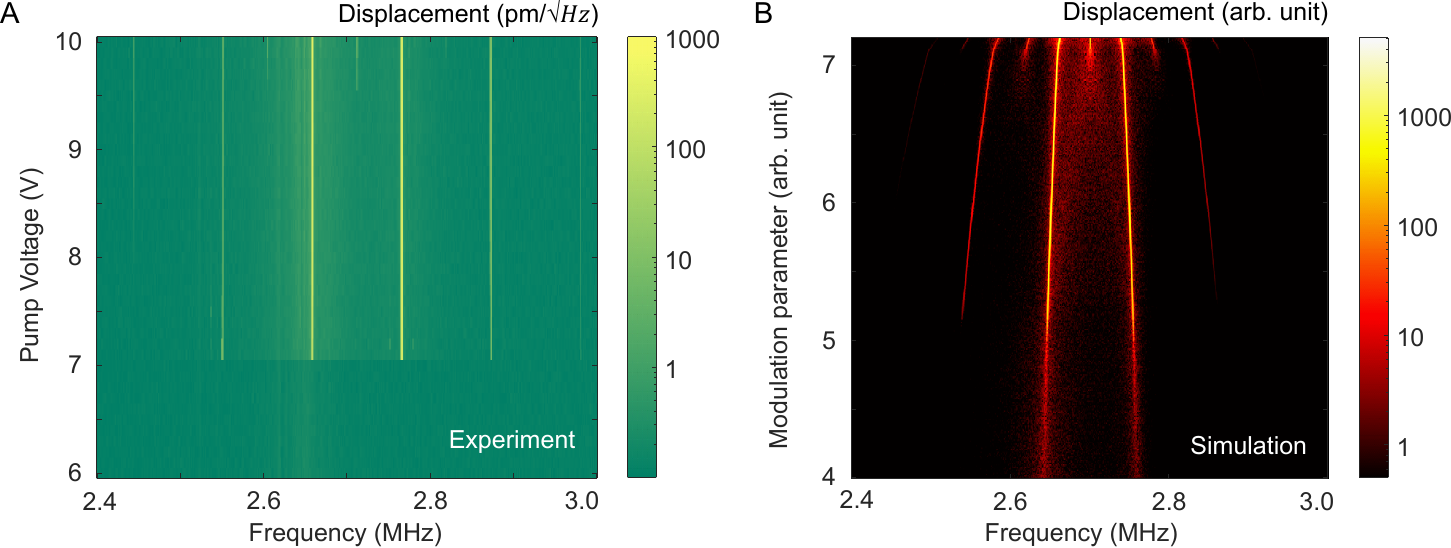}
	\caption{\textbf{Biharmonic phase:} \textbf{(A)} Experimental observation of biharmonic phase above a threshold value ($9.5 V$) of parametric pump voltage and keeping other parameters fixed. \textbf{(B)} Simulated spectrum shows similar behavior.}
	\label{biharmonic_2d} 
\end{figure}

%	Here,  $x_1$ and $x_2$ are the amplitudes of vertical displacements of graphene drums G1 and G2. $\gamma$, $m_g$, $h$, and $\alpha$ denotes linear damping coefficient, effective mass of graphene resonator, magnitude of the parametric drive and effective bilinear coupling between oscillators, respectively. $\omega$ is mean frequency of the oscillators and $\omega_p$ is parametric drive detuning. $\beta$ is the Duffing constant of the graphene resonator mode while $\eta$ represents the coefficient of nonlinear damping. We took $\chi_x = 0$ and $\chi_y= \pi /4$, phases of parametric drive, to simulate the system.
\vspace{5mm} 
\textbf{Biharmonic:} We observe transition from noncrystalline phase to anharmonic phase and anharmonic phase to biharmonic phase by changing pump voltage. First transition happens at threshold value $\sim 7 V$ and threshold value for biharmonic is $9.5 V$. We record frequency spectrum (see Fig. \ref{biharmonic_2d}A) and simulate similar result (Fig. \ref{biharmonic_2d}B) using coupled nonlinear equations (\ref{mean_field}).

%\begin{figure}
%	\centering
%	\includegraphics[scale=0.8]{SI_fig8_V4.pdf}
%	\caption{\textbf{Continuum:} \textbf{(A)} Cross-section of continuum observed in the frequency domain (see Fig.~4D in main text) corresponds to pump frequency value of $5.52$ MHz.  \textbf{(B)} Numerical result of two graphene model with an effective coupling shows similar kind of spectrum for pump strength value of 0.0651 (arb. unit) with a $\pi/4$ phase difference between parametric driving signals of the two drums.\textbf{(C)} Numerical observation of different time crystalline phases in the frequency domain for specific values of parameters with detuned pump frequency.} 
%	\label{continuum}	
%\end{figure}

\begin{figure}[H]
	\centering
	\includegraphics[scale=0.8]{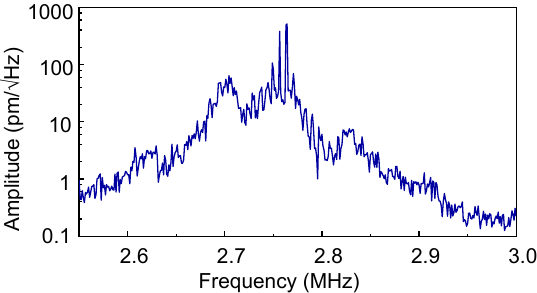}
	\caption{\textbf{Continuum:} Cross-section of continuum observed in the frequency domain (see Fig.~3A in main text) corresponds to pump frequency value of $5.52$ MHz.} 
	\label{continuum}	
\end{figure}

\vspace{5mm} 
%\textbf{Continuum}: Apart from three different time crystalline phases we observe another phase above the parametric pump threshold value which is not discussed in details in this work. For certain parameters, we observe amplification of a wide spectrum around half of the pump frequency (see Fig. \ref{continuum}A) which we identify as continuum. We observe same feature while simulating our coupled equation for very specific range of parameters (Fig. \ref{continuum}B). We are still investigating this phase to understand in details.
\textbf{Continuum}: In the experiment, apart from three different time crystalline phases we observe another phase above the parametric pump threshold value which is not discussed in detail in this work. For certain parameters, we observe amplification of a wide spectrum around half of the pump frequency (see Fig. \ref{continuum}) which we identify as continuum.

%We study these time crystalline phases experimentally by tuning pump frequency (Fig.~4D in main text). A numerical plot (Fig. \ref{continuum}C) with two coupled graphene mode shows few features same as Fig.~4D (main text) as also suggestive from Fig.~3 (main text) but experimental data is much more richer than Fig.~\ref{continuum}C due to many-body nature of the system, variation of number of modes, coupling and detuning  in experiment. 

\newpage
\subsection{Perturbative analysis for a mean-field phase diagram}
To understand analytically,	we introduce a small book-keeping parameter $\epsilon \equiv 1$ to keep track the order of perturbative equation. 
We define, $\omega_0\equiv\omega-\epsilon\Omega_p/2$. Now the model equations take the form,
\begin{subequations}
	\begin{eqnarray}
		\label{eq:redefined_x}
		&&\ddot{x}_1+\epsilon \gamma \dot{x}_1 + \left(\omega-\epsilon \frac{\delta\omega_1}{2}\right)^2 x_1+ \epsilon \frac{h}{m_g} \cos(2\omega t+\chi_1) x_1
		+\epsilon \frac{\eta}{m_g}x_1^2 \dot{x}_1 + \epsilon \frac{\beta}{m_g} x_1^3-\epsilon\frac{\alpha}{m_g} x_2=f_1(t)\\
		\label{eq:redefined_y}
		&&\ddot{x}_2+\epsilon \gamma \dot{x}_2 + \left(\omega-\epsilon \frac{\delta\omega_2}{2}\right)^2 x_2+ \epsilon \frac{h}{m_g}  \cos(2\omega t+\chi_2) x_2 +\epsilon \frac{\eta}{m_g}x_2^2 \dot{x}_2 + \epsilon \frac{\beta}{m_g} x_2^3-\epsilon\frac{\alpha}{m_g} x_1=f_2(t),
	\end{eqnarray}
\end{subequations}

where, $\delta\omega_1=\Omega_p+\Delta$ and $\delta\omega_2=\Omega_p-\Delta$.

We are now interested in the robust solutions of the above equations. The system is analytically tractable only through perturbative analysis. Considering only the deterministic part, we adopt the multiple time scales method (any equivalent perturbative techniques could be used) and start by writing the ansatz for the  equation~(\ref{eq:redefined_x}) and equation~(\ref{eq:redefined_y}) as,
\begin{subequations}
	\begin{eqnarray}
		\label{eq:ansatz_x}
		&& x_1(t,T)=x_1^0(t,T)+\epsilon x_1^1(t,T)+\cdots\\
		\label{eq:ansatz_y}
		&& x_2(t,T)=x_2^0(t,T)+\epsilon x_2^1(t,T)+\cdots
	\end{eqnarray}
\end{subequations}
As here we have two time scales $t$ and $T\equiv\epsilon t$, therefore Using the ansatz given by equations~(\ref{eq:ansatz_x})-(\ref{eq:ansatz_y}) in equations~(\ref{eq:redefined_x})-(\ref{eq:redefined_y}), the 
$0^{\textrm{th}}$ order and the $1^{\textrm{st}}$ order equations  in $\epsilon$ are,
\begin{subequations}
	\begin{eqnarray}
		\label{eq:x0}
		&&D_{tt}x_1^0+\omega^2 x_1^0=0,\\
		\label{eq:y0}
		&&D_{tt}x_2^0+\omega^2 x_2^0=0,\\
		\label{eq:x1}
		&&D_{tt}x_1^1+\omega^2 x_1^1=-2(D_tD_Tx_1^0)-\gamma (D_tx_1^0)+\omega\delta\omega_1 x_1^0-\frac{h}{m_g}\cos(2\omega t+\chi_1)x_1^0-\frac{\eta}{m_g} {(x_1^0)}^2(D_t x_1^0)\nonumber\\
		&&-\frac{\beta}{m_g}{(x_1^0)}^3+\frac{\alpha}{m_g}x_2^0,\\
		\label{eq:y1}
		&&D_{tt}x_2^1+\omega^2 x_2^1=-2(D_tD_T x_2^0)-\gamma (D_t x_2^0)+\omega\delta\omega_2 x_2^0-\frac{h}{m_g}\cos(2\omega t+\chi_2)x_2^0-\frac{\eta}{m_g}{(x_2^0)}^2(D_t x_2^0)\nonumber\\
		&&-\frac{\beta}{m_g}{(x_2^0)}^3+\frac{\alpha}{m_g}x_1^0.
	\end{eqnarray}
\end{subequations}
The solution of the $0^{\textrm{th}}$ order equations for $x_1$ and $x_2$ are,
\begin{subequations}
	\begin{eqnarray}
		\label{eq:sol_x0}
		&&x_1^0=\left(\frac{a(T)}{2} \exp(i \omega t) + \textrm{c.c.}\right),\\
		\label{eq:sol_y0}
		&&x_2^0=\left(\frac{b(T)}{2} \exp(i \omega t) + \textrm{c.c.}\right).
	\end{eqnarray}	
\end{subequations}
Using the zeroth order solutions in Eq.~(\ref{eq:x1}) and Eq.~(\ref{eq:y1}) and equating the coefficients of $\exp(i\omega t)$ equal to zero, to remove the secular terms, in differential equations for $x_1$ and $x_2$, we get the
evolution of the real and the imaginary parts of the amplitudes, given by,
\begin{subequations}
	\begin{eqnarray}
		\label{eq:dar_dt}
		&\frac{da_r}{dT}=&-\frac{\gamma}{2}a_r - \frac{h}{4 \omega m_g}(a_r \sin \chi_1 - a_i \cos \chi_1) - \frac{3\beta}
		{8 \omega m_g}(a_r^2+a_i^2)a_i-\frac{\eta}{8 m_g}(a_r^2+a_i^2)a_r+\frac{\alpha}{2\omega m_g}b_i +\frac{\Delta}{2}a_i\nonumber\\
		&&+\frac{\Omega_p}{2}a_i,\\
		&\frac{db_r}{dT}=&-\frac{\gamma}{2}b_r - \frac{h}{4 \omega m_g}(b_r \sin \chi_2 - b_i \cos \chi_2) - \frac{3\beta}
		{8 \omega m_g}(b_r^2+b_i^2)b_i-\frac{\eta}{8 m_g}(b_r^2+b_i^2)b_r+\frac{\alpha}{2\omega m_g}a_i -\frac{\Delta}{2}b_i\nonumber\\
		&&+\frac{\Omega_p}{2}b_i,\\
		&\frac{da_i}{dT}=&-\frac{\gamma}{2}a_i + \frac{h}{4 \omega m_g}(a_r \cos \chi_1 + a_i \sin \chi_1) + \frac{3\beta}
		{8 \omega m_g}(a_r^2+a_i^2)a_r-\frac{\eta}{8 m_g}(a_r^2+a_i^2)a_i-\frac{\alpha}{2\omega m_g}b_r -\frac{\Delta}{2}a_r\nonumber\\
		&&-\frac{\Omega_p}{2}a_r,\\
		\label{eq:dbi_dt}
		&\frac{db_i}{dT}=&-\frac{\gamma}{2}b_i + \frac{h}{4 \omega m_g}(b_r \cos \chi_2 + b_i \sin \chi_2) + \frac{3\beta}
		{8 \omega m_g}(b_r^2+b_i^2)b_r-\frac{\eta}{8 m_g}(b_r^2+b_i^2)b_i-\frac{\alpha}{2\omega m_g}a_r +\frac{\Delta}{2}b_r\nonumber\\
		&&-\frac{\Omega_p}{2}b_r,
	\end{eqnarray}
\end{subequations}
where, $a_r$ and $b_r$ are real parts of $a$ and $b$ respectively.  $a_i$ and $b_i$ are the imaginary parts of $a$ and $b$ respectively.

Till now, we have discussed the theoretical model and corresponding equations of motion. Using the multiple timescales analysis, we find the first order differential equations for real and imaginary parts of amplitudes in equations~(\ref{eq:dar_dt})-(\ref{eq:dbi_dt}), which forms the basis to study the dynamics of the system in four dimensional phase space via linear stability analysis(see Fixed point analysis of appendix in main text).

\newpage
\section{Quintessential many-body nature of phases: fluctuation and rigidity}

An interacting many-body system shows different time crystalline phases depending on coupling parameter, drive strength and fluctuation present in the system. Due to high mode density, graphene peak interacts with multiple $SiNx$ modes and number depends on the parameters of the experiment.

\subsection{Mode density of $SiNx$ resonator}

We do a rough estimation of $SiNx$ mode density of the sample we are working on. Using a lock-in amplifier we send a signal to drive the system weakly. Tuning frequency around the experiment range ($2$ $MHz$ to $4$ $MHz$) we estimate average mode density $88$ peaks/$MHz$. Now we can roughly estimate number of modes showing different time crystalline phases. We will consider modes having frequency in between two graphene drums. 

For subharmonic response, frequency difference between graphene peaks are $\sim 100$ $kHz$ and full width of half maxima (FWHM) of a graphene peak is $34.42$ $kHz$. So we have $\sim 12$ $SiNx$ modes in the frequency span of $134.42$ $KHz$. For anharmonic and biharmonic case, difference between graphene peaks are $150$ $KHz$ and $125$ $kHz$ respectively. So, we have $\sim 16$ and $\sim 14$ $SiNx$ modes close to graphene drums when anharmonic and biharmonic phases emerged.   
\begin{figure}[H]
	\centering
	\includegraphics[scale=0.5]{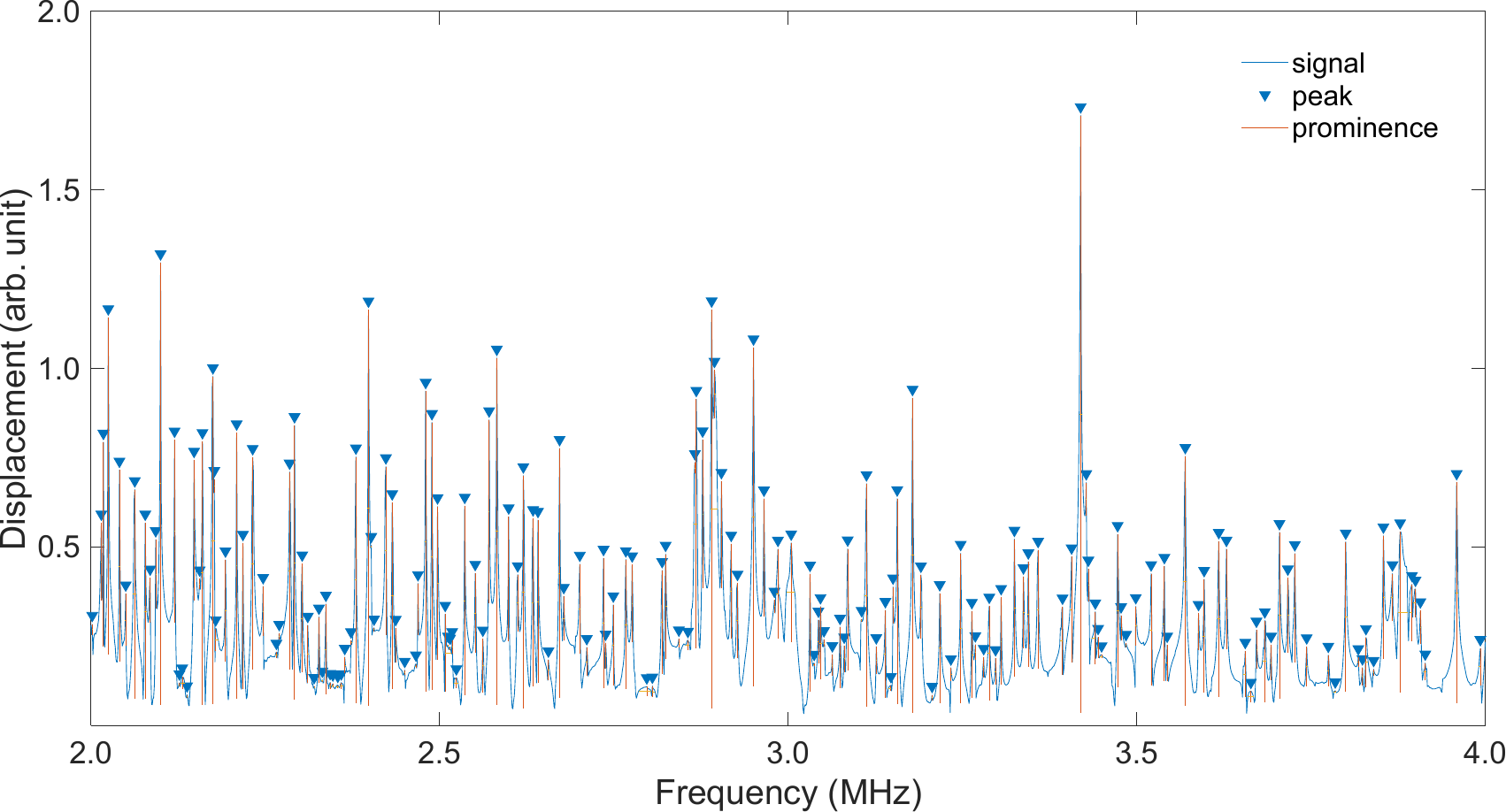}
	\caption{\textbf{ $SiNx$ mode density}}
	\label{SiNx_mode_density} 
\end{figure}

\subsection{Nature of added noise}
\begin{figure}
	\centering
	\includegraphics[scale=0.25]{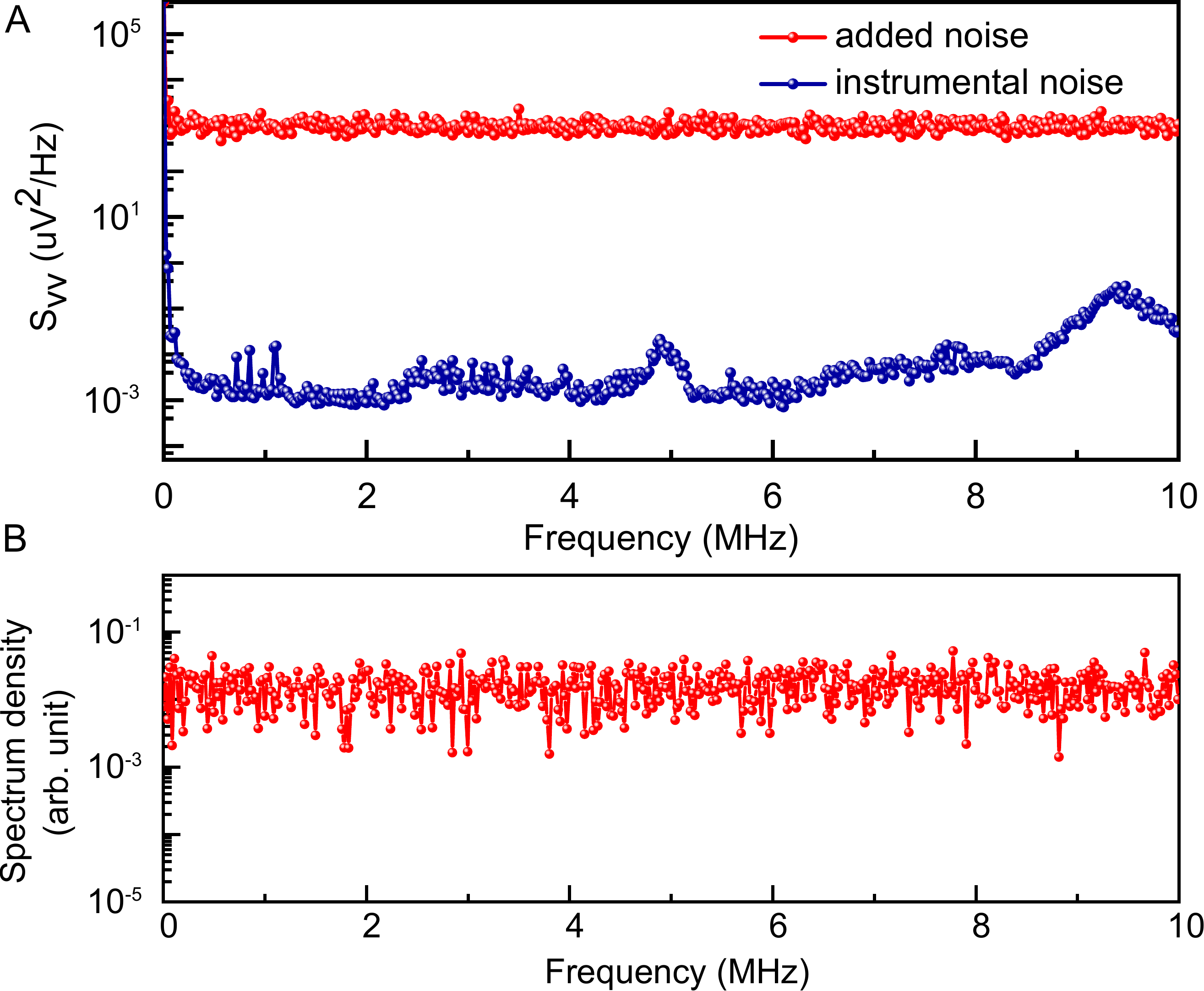}
	\caption{\textbf{Flat noise spectrum:}\textbf{(A)} Experimental data for added white noise (red) and instrumental noise (blue). \textbf{(B)} White noise spectrum of numerical random function.} 
	\label{flat_spectrum}
\end{figure}

To test rigidity of time crystalline phases we feed in noise across the capacitive arrangement and record the spectrum with increasing r.m.s. amplitude of the noise ($V_{r.m.s.}$). Experimentally noise is added using a function generator. To understand the nature of noise, we look at Fourier transform of added noise.  Blue curve in Fig. \ref{flat_spectrum}A is instrumental noise floor without any added noise for a large range of frequency (0 to 10 MHz). Added noise gives a flat spectrum (red curve in Fig. \ref{flat_spectrum}A) for the same range.

For numerical analysis, we choose a random function to add noise to the system. Amplitude of the function is Gaussian for uniform time step. Fourier transform of this function also gives a flat spectrum. So we will use this function for further numerical analysis.\\

\subsection{Additive and multiplicative term from capacitive arrangement}
Our system is kept in a capacitive arrangement and force is applied through metallic contacts of sample and back gate. For simplicity of calculation, we rewrite equation \ref{nl_HO} for $\alpha=0$, adding linear damping ($\gamma$) and a force term $F(x,t)$:
\begin{equation}\label{EOM}
	\ddot{x}+ \gamma \dot{x} + \omega_0^2 x +  \frac{\beta}{m} x^3=F(x,t).
\end{equation}
The total force $F(x,t)$ acting on the sample due to the applied voltage $V$ to capacitor combines an elastic term and an electrostatic term. For a time varying voltage, we can write
$V(t)=V_{DC} + V_{AC}(t)+ V_{noise}(t)$, where $V_{AC}$ and $V_{noise}(t)$ represent parametric pumping and externally added noise respectively.

From stored elastic energy of the system \cite{Chen}, we calculate the elastic force to be:
\begin{equation}
	F_{el}= -\frac{16 E A \chi_0 }{3 L} x - \frac{256 E A}{9 L^3} x^3.
\end{equation}
 
Here, $E,A,\chi_0$ and $L$ represent young modulus, cross sectional area, built in strain and diameter of the suspended material respectively.

Aditionally, for a parallel plate capacitor, we can write 

\begin{equation}
	F_{es}= -\frac{\epsilon_0 A V(t)^2}{2(d_0-x)^2}. 
\end{equation}

where, $\epsilon_0$ and $d_0$ represent permittivity of air and mean distance between parallel plates respectively.

By expanding $F_{es}$ around $x_0$ up to first order \cite{lardies:hal-02300589}, we get the total force acting on the sample perpendicular to the surface:
\begin{equation}
	F= F_{el}-F_{es}= -\frac{256 E A}{9 L^3} x^3 + \left( \frac{\epsilon_0 A V(t)^2}{(d_0-x_0)^3}-\frac{16 E A \chi_0 }{3 L}\right) x +\frac{\epsilon_0 A V(t)^2}{2(d_0-x_0)^2}-\frac{\epsilon_0 A V(t)^2}{(d_0-x_0)^3} x_0.
\end{equation}

This leads to the following equation of motion:

\begin{equation} \label{EOM2}
	\ddot{x}+ \gamma \dot{x} + \left(\omega_0^2 + \frac{16 E A \chi_0 }{3 L}- \frac{\epsilon_0 A V(t)^2}{(d_0-x_0)^3}\right) x +  \left(\frac{\beta}{m}+\frac{256 E A}{9 L^3} \right) x^3 = \frac{\epsilon_0 A V(t)^2 (d_0-3 x_0)}{2(d_0-x_0)^3}.
\end{equation}
%\begin{figure}
%	\centering
%	\includegraphics[scale=0.6]{additive_mult_noise_V1.png}
%	\caption{\textbf{Effect of noise:} \textbf{(A)} Lorentzian fit on noise spectrum for increasing (from blue to yellow) values of added noise. \textbf{(B)} Change in noise floor with increasing additive noise strength.  \textbf{(C)} Lorentzian fit on noise spectrum for increasing (from blue to yellow) values of multiplicative noise. \textbf{(D)} Frequency shift with increasing noise strength.}
%	\label{additive_mult_noise}
%\end{figure}

So the external noise fed through capacitive arrangement has two contributions to the equation of motion. Multiplicative noise term modifies the frequency  and  additive noise term acts as a direct drive while the former being $\frac{d_0}{x}$ order of magnitude higher.

\subsection{Noise axis calibration}

\begin{figure}[H]
	\centering
	\includegraphics[scale=0.6]{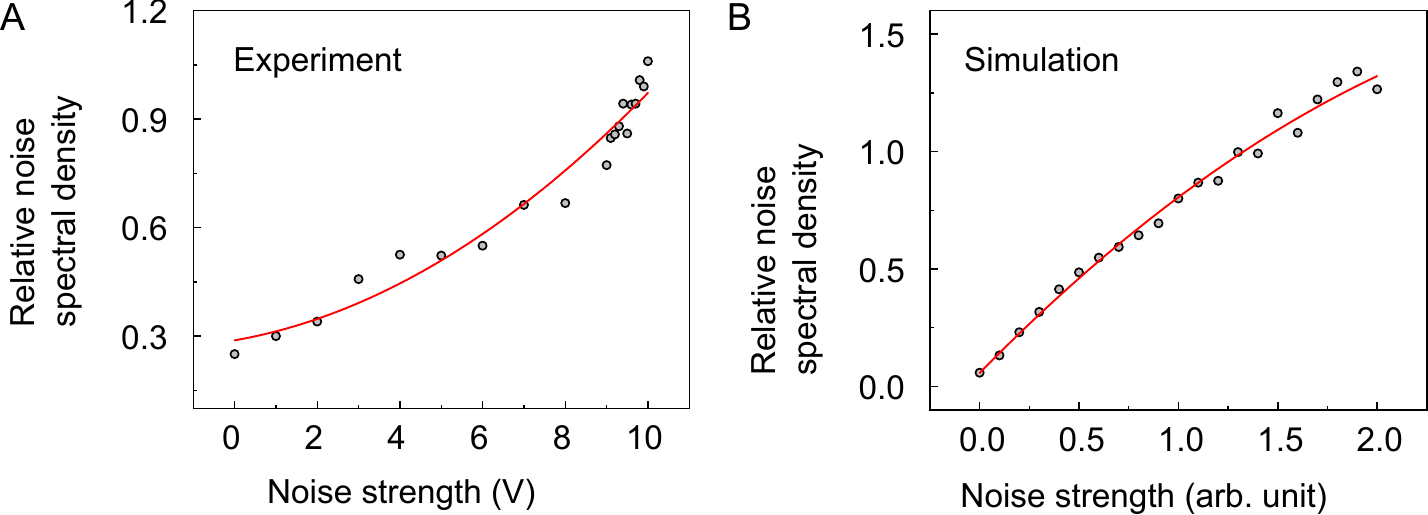}
	\caption{\textbf{Noise induced phase transition:}\textbf{(A)} Experimental and \textbf{(B)} numerical relative noise spectral density vs input noise strength is plotted.}
	\label{robust_compare} 
\end{figure}

To compare experimental observation with numerical result, noise axis is calibrated by measuring the change in noise floor with additive noise. In experiments, the noise floor (extracted via Lorentzian fits) increases steadily with increasing external noise strength. Numerically we observe similar behaviour if we consider the external noise to be additive only, which is consistent with equation \ref{EOM2}. To make this comparison dimensionless, we define a new parameter, relative noise spectral density, by taking the ratio of the fitted noise floor to the thermal peak amplitude. This unitless quantity serves as a convenient control parameter for analyzing noise-induced phase transitions in both experiment and simulation (Fig. \ref{robust_compare}).

\section{Robustness to fluctuations}

\begin{figure}[H]
	\centering
	\includegraphics[scale=0.6]{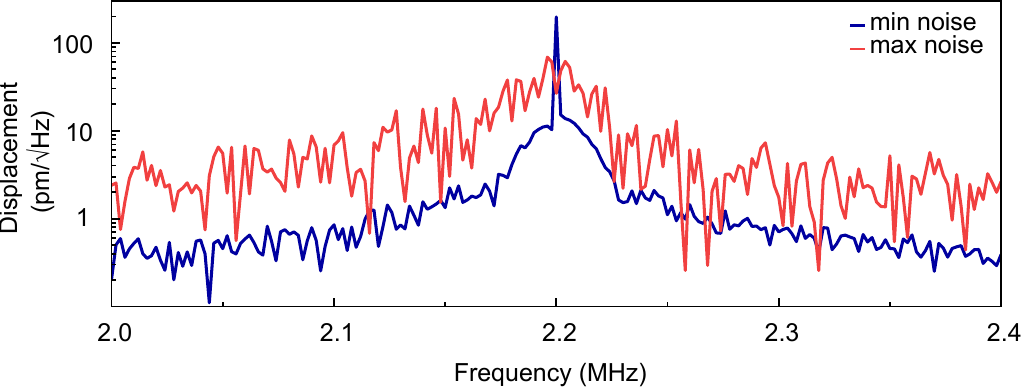}
	\caption{\textbf{Robustness of subharmonic phase:} Numerical displacement spectral density shows subharmonic response (blue) gets destroyed slowly with increasing noise strength (red).} 
	\label{robust_exp}
\end{figure}

We focus on subharmonic phase to study robustness of the phase against added noise.  In experiment the phase remains stable for certain threshold value of relative noise spectral density (Fig. 3B in main text). In numerical analysis, we observe similar nature of spectra (Fig. \ref{robust_exp}) but do not observe sharp transition in the two-mode model (Fig. 7 in main text).

\bibliographystyle{apsrev4-2} 
\bibliography{reference_bibtex_V9}

\end{document}